# Unveiling the nucleation and growth of Zr oxide precipitates in internally oxidized $Nb_3Sn$ superconductors

Jaeyel Lee[1], Zugang Mao[2], Dieter Isheim[2,3], David N Seidman[2,3], Xingchen Xu[1]

[1] *Fermi National Accelerator Laboratory, Batavia, IL, 60510, USA*

[2] *Department of Materials Science and Engineering, Northwestern University, Evanston, IL, 60208, USA*

[3] *Northwestern University Center for Atom-Probe Tomography (NUCAPT), Evanston, IL, 60208, USA*

**Abstract**

We report on atomic-scale analyses of nucleation and growth of Zr oxide precipitates and the microstructural evolution of internally oxidized $Nb_3Sn$ wires for high-field superconducting magnet applications, utilizing atom probe tomography (APT), transmission electron microscopy (TEM), and first-principles calculations. APT analyses reveal that oxygen and zirconium are already segregated at grain boundaries (GBs) in the unreacted Nb-1Zr-4Ta (at.%) alloy prior to forming $Nb_3Sn$ through reacting the Nb alloy with Sn and $SnO_2$. After forming $Nb_3Sn$, Zr oxide precipitates nucleate both at the $Nb_3Sn$/Nb heterophase interfaces and in the $Nb_3Sn$ grains, driven by the small solubilities of Zr and O in $Nb_3Sn$ compared to their value in Nb. A high number density ($N_v$) of Zr oxide nanoprecipitates is observed in the $Nb_3Sn$ layers, ~$10^{23}$ $m^{-3}$, with a mean diameter <10 nm for a heat treatment at 625 °C. Quantitative APT and TEM analyses of the Zr oxide precipitates in the reacted $Nb_3Sn$ layers elucidate details of the nucleation, growth, and coarsening processes of the Zr oxide precipitates in $Nb_3Sn$. First-principles calculations and classical nucleation theory are employed to study the nucleation of Zr oxide precipitates in $Nb_3Sn$ and to estimate the maximum energy barrier and critical radius for nucleation. Our research unveils the kinetic pathways for nucleation and growth of Zr oxide precipitates and the microstructural evolution of $Nb_3Sn$ layers, which helps to understand and improve the superconducting properties of internally oxidized $Nb_3Sn$ wires for use in high-field superconducting magnets.

## 1. Introduction

$Nb_3Sn$ is a superconducting material with a critical temperature $T_c$ = 18.3 K and important applications in high-field magnets. These include solenoid magnets up to 23.5 T for nuclear magnetic resonance (NMR) spectrometers and other commercial magnets [1]. $Nb_3Sn$ is also being used for large-scale science projects, e.g., for the Toroidal Field (TF) and the Central Solenoid (CS) coils for the International Thermonuclear Experimental Reactor (ITER) project [2] as well as the interaction region quadrupole magnets for the ongoing High-Luminosity upgrade to the Large Hadron Collider (LHC) [3]. It is also the baseline superconductor for the accelerator magnets for the Future Circular Collider (FCC)-hh, which is proposed to succeed the LHC to explore new particle physics [4, 5]. However, the performance of present state-of-the-art commercial $Nb_3Sn$ superconductor materials still cannot meet the requirements of FCC-hh magnets [5]. Specifically, the critical current density ($J_c$) had been at a plateau significantly below the FCC $J_c$ specification for two decades [6].

To further improve $Nb_3Sn$'s $J_c$ value Xu et al. developed a new type of $Nb_3Sn$ superconducting wire based on internal oxidation with improved magnetic flux pinning [7, 8]. After several years of development, its $J_c$ has met the FCC $J_c$ specification [9]. Furthermore, it also displays a series of interesting important properties, which are distinct from conventional $Nb_3Sn$ materials, such as a flatter $J_c$-$B$ curve and reduced low-field magnetization [10]. A $Nb_3Sn$ wire is composed of a number of $Nb_3Sn$ subelements embedded in a Cu matrix, see **Fig. 1(a)**. A conventional $Nb_3Sn$ subelement is synthesized from precursor materials, including Sn, Cu and a Nb alloy with a heat treatment to convert the precursors to $Nb_3Sn$ with some Nb remaining to protect the Cu matrix. Copper is added to promote the formation of $Nb_3Sn$ during a heat treatment [11] and the most commonly used Nb precursor alloy is Nb-7.5wt.%Ta, where Ta serves as a dopant to improve $Nb_3Sn$'s upper critical field $B_{c2}$ [12]. By comparison, in an internal-oxidation-type $Nb_3Sn$ subelement, O is added (e.g., by adding $SnO_2$ to the Sn and Cu), and a Nb-Ta-Zr (or Nb-Ta-Hf) precursor alloy is utilized so that O selectively oxidizes Zr (or Hf) to form oxide nanoprecipitates in $Nb_3Sn$ during a heat treatment owing to the much higher affinity of Zr (or Hf) for O than that of Nb for O [7, 13-17]. The oxide nanoprecipitates increase the $J_c$ of $Nb_3Sn$ superconductors in two ways: (1) they can dramatically refine the $Nb_3Sn$ grain diameter through the Zener grain boundary pinning effect, and a smaller grain diameter leads to an increased $J_c$ as grain boundaries (GBs) are magnetic flux pinning centers; (2) these nanoprecipitates themselves have the right diameter to serve as flux pinning centers and thereby improve $J_c$ [9]. Hence, oxide nanoprecipitates play a central role in improving the superconducting performance of $Nb_3Sn$ conductors.

To date, a systematic study of Zr oxide nanoprecipitates in internally oxidized $Nb_3Sn$ is absent, specifically nanoprecipitate diameter distribution, number density per unit volume, and their evolution with heat treatment. These quantities are critical for the performance of $Nb_3Sn$ superconductors because they determine directly the effectiveness of nanoprecipitates as flux pinning centers. Nanoprecipitates approaching twice the coherence length of a superconductor (~3.5 nm for $Nb_3Sn$ at 4.2 K) are the most efficient for flux pinning, and a high number density of nanoprecipitates is critical for increasing the pinning force. It is necessary to understand what factors determine the diameters and number densities of Zr oxide nanoprecipitates, so that we can design the processing of $Nb_3Sn$ superconductors to generate nanoprecipitates with the optimal diameters and highest number densities. To accomplish these results we need to understand how Zr oxide nanoprecipitates nucleate and grow in $Nb_3Sn$.

To answer these questions, we studied different regions in $Nb_3Sn$ filaments, including unreacted Nb alloy, $Nb_3Sn$/Nb interfaces, and $Nb_3Sn$ layers, employing atom-probe tomography (APT) and transmission electron microscopy (TEM). We also explored the kinetics and thermodynamics for Zr oxide precipitation using first-principles calculations. The results demonstrate that the $Nb_3Sn$/Nb interfaces play critical roles in the nucleation and growth of Zr oxide nanoprecipitates. The effects of the microstructure of internally oxidized $Nb_3Sn$ wires on their superconducting properties are also discussed.

## 2. Experimental procedure and theoretical details

### 2.1. Experimental procedure

The internal-oxidation-type $Nb_3Sn$ wire used in this work was fabricated at Hyper Tech Research Inc. based on the powder-in-tube method, with details described in refs. [6, 9, 18, 19]. Each $Nb_3Sn$ filament is fabricated by filling a mixture of Sn, $SnO_2$, and Cu powders into a Nb-1 at.% Zr-4 at.% Ta tube. The $SnO_2$ amount was such that the O concentration was more than sufficient to oxidize all the Zr. The Nb alloy was in excess relative to Sn so after a full reaction between Sn and Nb, there is a residual Nb layer. Heat treatments at 625 °C for 740 h and at 700 °C for 120 h were employed to study the influence of heat treatment temperature on the evolution of Zr oxide nanoprecipitates in $Nb_3Sn$.

The microstructures of the $Nb_3Sn$ samples were systematically analyzed utilizing scanning electron microscopy (SEM), S/TEM, and APT. SEM imaging and sample preparations for TEM foils and APT nanotips were performed using a FEI Helios Nanolab 600 focused ion beam (FIB). TEM foils and APT nanotips were thinned by Ga ions at 30 kV and 87 pA to 0.4 nA: the surface damage was removed by Ga ions at 5 kV and 47 pA. High-resolution STEM imaging of the $Nb_3Sn$ samples was performed employing an aberration-corrected JEOL ARM 200. For APT analyses, a Cameca (Madison, WI) LEAP5000 XS was utilized. A picosecond ultraviolet laser (wavelength =

355 nm) was employed to dissect nanotips utilizing 30 pJ laser pules, at a pulse repetition rate of 500 kHz: the detection rate was 0.5-1.0 %. 3D reconstructions of nanotips were performed using IVAS 3.8.1 (Cameca, Madison, WI). Each peak in an APT mass spectrum was carefully analyzed and identified. Details of the identification of each peak in a mass spectrum are presented in **Fig. S1** of the supplementary material section.

### 2.2. Theoretical details

Thermodynamics of nucleation of Zr oxide precipitates in $Nb_3Sn$ are studied using classical nucleation theory (CNT) and first-principles calculations from which the bulk energies of $ZrO_2$ and $Nb_3Sn$, lattice strains, and interfacial energies are determined. The first-principles calculations employ the plane wave pseudopotential total energy method as implemented in the *Vienna ab initio simulation package* (VASP). We employ projector augmented wave (PAW) potentials [20, 21] and the generalized gradient approximation (GGA) of the Perdew-Burke-Ernzerhof (PBE) methodology for the exchange-correlation energy [22]. All structures are fully relaxed with respect to volume, as well as all cell-internal atomic coordinates. We carefully consider the convergence of results with respect to energy cutoff and k-points. A plane-wave basis set is used with an energy cutoff of 750 eV to represent the Kohn-Sham wave functions. The summation over the Brillouin zone for the bulk structures is performed on a 12×12×12 Monkhorst-pack k-point mesh for all calculations. The magnetic spin-polarized methodology is applied in all the calculations. The calculated lattice parameter of $Nb_3Sn$ is 5.332 Å, which is in agreement with experimental results, 5.289 Å for $Nb_3Sn$, with a formation energy of -0.1635 eV/atom. The calculated lattice parameters of monoclinic $ZrO_2$ are a = 5.266 Å, b = 5.235 Å, and c = 5.416 Å, with a formation energy of -3.822 eV/atom. The calculated lattice parameters of tetragonal $ZrO_2$ are a = 3.643 Å, and c = 5.318 Å with a formation energy of -3.779 eV/atom. The calculated lattice parameter of cubic $ZrO_2$ is a

= 3.642 Å with a formation energy of -3.752 eV/atom. The interfacial structure of $ZrO_2$ (110) /$Nb_3Sn$ (110) is built employing eight layers of atoms for both structures, which are fully relaxed with an 8×2×2 k-point mesh.

## 3. Experimental and theoretical results

### 3.1. The microstructures of $Nb_3Sn$ layers heat treated at 625 °C and 700 °C

An SEM image of a cross-section of a $Nb_3Sn$ wire after reaction at 625 °C for 740 h is displayed in **Fig. 1(a).** The concentric layered structure inside an individual filament of the wire I shown in **Fig. 1(b)** with Cu, residual Nb alloy, $Nb_3Sn$/Nb reaction front interface, formed $Nb_3Sn$ layer, and a core with residual $SnO_2$. The yellow rectangles and dots mark the regions for detailed APT analysis. High-angle annular dark-field (HAADF)-STEM images of $Nb_3Sn$ layers after inter-reaction heat-treatments at 625°C for 740 h and at 700 °C for 120 h are displayed in **Fig. 1(c,d)**, respectively, with the $Nb_3Sn$/Nb reaction front interface at the far right-hand side. These images display gradual changes in the grain diameters of the $Nb_3Sn$ layers at different distances from the $Nb_3Sn$/Nb interfaces in both samples. For the $Nb_3Sn$ sample heat-treated at 625 °C, $Nb_3Sn$ grains directly at the $Nb_3Sn$/Nb interface have an average grain diameter of 36±5 nm. The $Nb_3Sn$ grains located further away from the interface gradually grow after the $Nb_3Sn$/Nb inter-reaction front has passed through, with the grain size almost doubling to 70 ±12 nm at a distance of ~5 μm from the $Nb_3Sn$/Nb interface, **Fig. 1(e)**. For the wire heat-treated at 700 °C, $Nb_3Sn$ grains directly at the interface have an average grain diameter of 48±6 nm. These grains continue to grow and coarsen, resulting in average grain diameters of 81±13 nm and 100±19 nm at a distance of 5 μm and 10 μm from the $Nb_3Sn$/Nb interface, respectively.

Notably, the average grain diameter throughout the entire $Nb_3Sn$ layer after heat treatment at 625 °C for 740 h is still less than ~70 nm, which is significantly smaller than that of comparable $Nb_3Sn$ superconductors without oxide precipitates, about 110 nm, for heat treatment at 625 °C [6]. In the case of $Nb_3Sn$ reacted at 700 °C for 120 h, there is moderate, but still limited grain growth, where the average grain diameter throughout the entire $Nb_3Sn$ layer is ~100 nm. This is the effect of grain boundary Zener pinning by Zr oxide nanoprecipitates on $Nb_3Sn$ GBs: see Sections 3.4 and 4 for additional details [23].

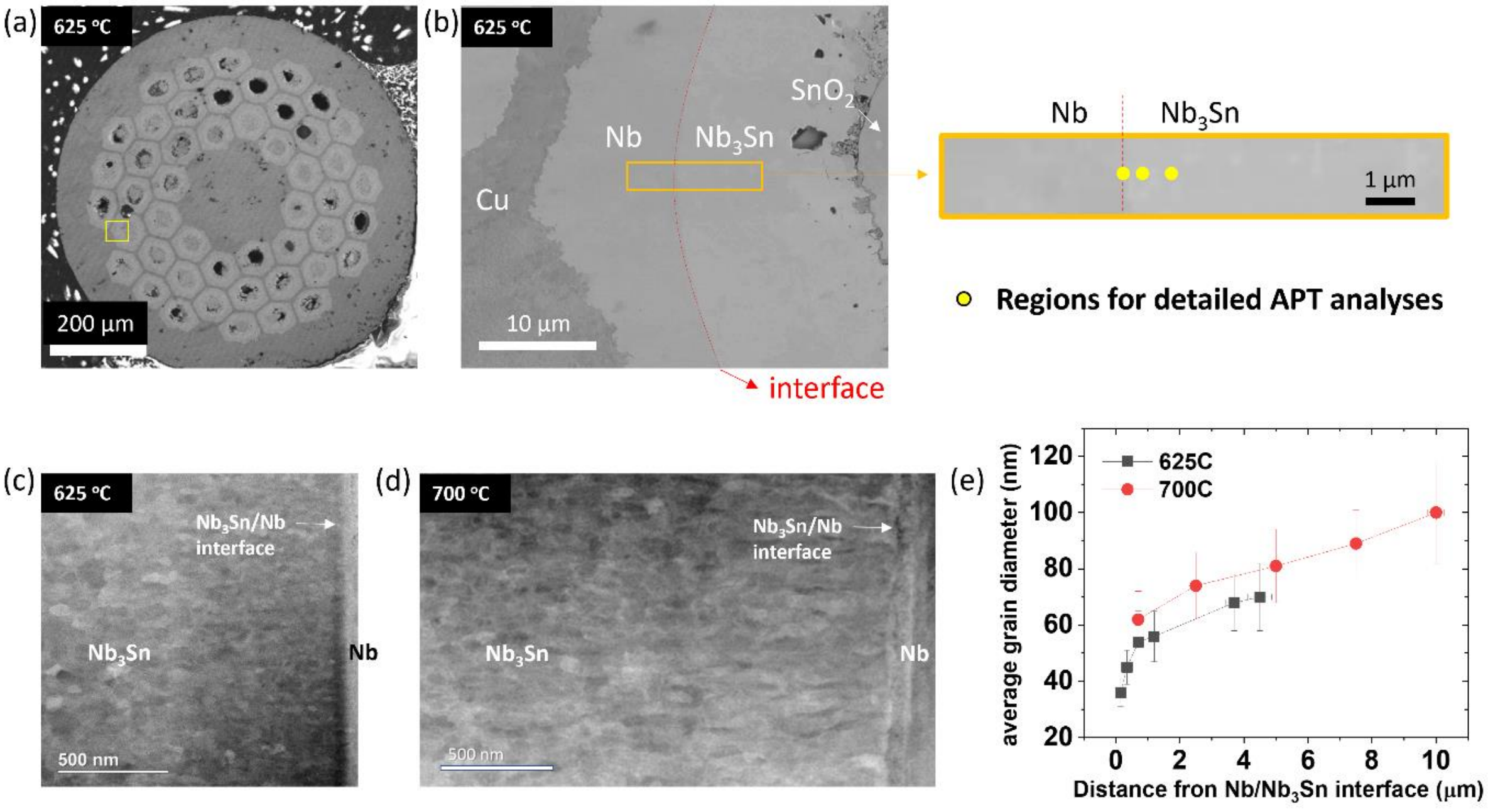

**Figure 1**. (a) Cross-sectional SEM images of $Nb_3Sn$ wires reacted at 625 °C for 740 h resolving individual filaments; (b) details of the layered structure within an individual filament with concentric Cu, Nb, $Nb_3Sn$, and $SnO_2$ layers as labeled, and regions for detailed APT analyses are marked with solid yellow circles. HAADF-STEM images of $Nb_3Sn$ layers with the Nb/$Nb_3Sn$ inter-reaction interface visible at the far right-hand side: (c) 625 °C for 740 h; and (d) 700 °C for 120 h heat treatments. (e) The plot of average grain diameters of $Nb_3Sn$ for 625 °C for 740 h and 700 °C for 120 h heat treatments, as a function of distances from the Nb/$Nb_3Sn$ interfaces, illustrates the growth of $Nb_3Sn$ grains during heat treatments. More significant growth of $Nb_3Sn$ grains is seen at 700 °C than at 625 °C.

### 3.2 Atom-Probe Tomographic analyses of the unreacted Nb alloy

In an earlier study, we determined that during the inter-reaction heat treatment, the $SnO_2$ is reduced to metallic Sn and the oxygen diffuses into the Nb alloy and dissolves in it, prior to the formation of $Nb_3Sn$ [24]. Thus, there is a possibility that Zr oxide nanoprecipitates form already in the Nb-Zr-Ta alloy before the interfacial reaction at Nb/$Nb_3Sn$ during heat treatments. To explore this possibility, we performed APT analyses of a region in the unreacted Nb alloy near the reaction interface of the sample heat-treated at 625 °C for 740 h: the results are displayed in **Fig. 2**. APT analyses show that the atomic concentrations of Zr and O in the Nb-Zr-Ta alloy are 1.0±0.1 at.% Zr and 5.5±0.2 at.% O, respectively.  3-D APT reconstructions of the positions of molecular ions (Zr-O, Nb-O), most likely representing Zr-O or Nb-O dimers in the alloy, and atomic ions (Nb, Zr, O) in the Nb nanotip are presented in **Figs. 2(a-d)**. There isn't clear evidence of crystalline Zr oxide or Nb oxide precipitates in the unreacted Nb alloy based on our TEM and APT analyses. However, the APT analyses indicate that Zr atoms bind to O and form Zr-O dimers in the Nb grains interior and GBs, **Fig. 2**. Additionally, O atoms are segregated at GBs; the atomic concentration of O in the Nb alloy (5.5±0.2 at.% O) is greater than the solubility of O in Nb at 625 $^{o}$C, ~2 at.% O [25]. This leads to high concentrations of Zr-O and Nb-O dimers at Nb GBs. The atomic concentration profiles across a Nb GB, **Fig. 2 (e)**, illustrate that the Gibbs interfacial excesses of Zr and O at the GB are ~4 atoms/nm$^2$ and ~18 atoms/nm$^2$, respectively. The 3D APT reconstruction of a nanotip of Nb establishes that although GBs are preferential sites for the segregation of Zr-O and Nb-O clusters, there are no discrete Zr oxide precipitates formed in the Nb alloy yet. To search for nucleation sites for $ZrO_2$ nanoprecipitates, we studied $Nb_3Sn$/Nb inter-reaction interfaces next.

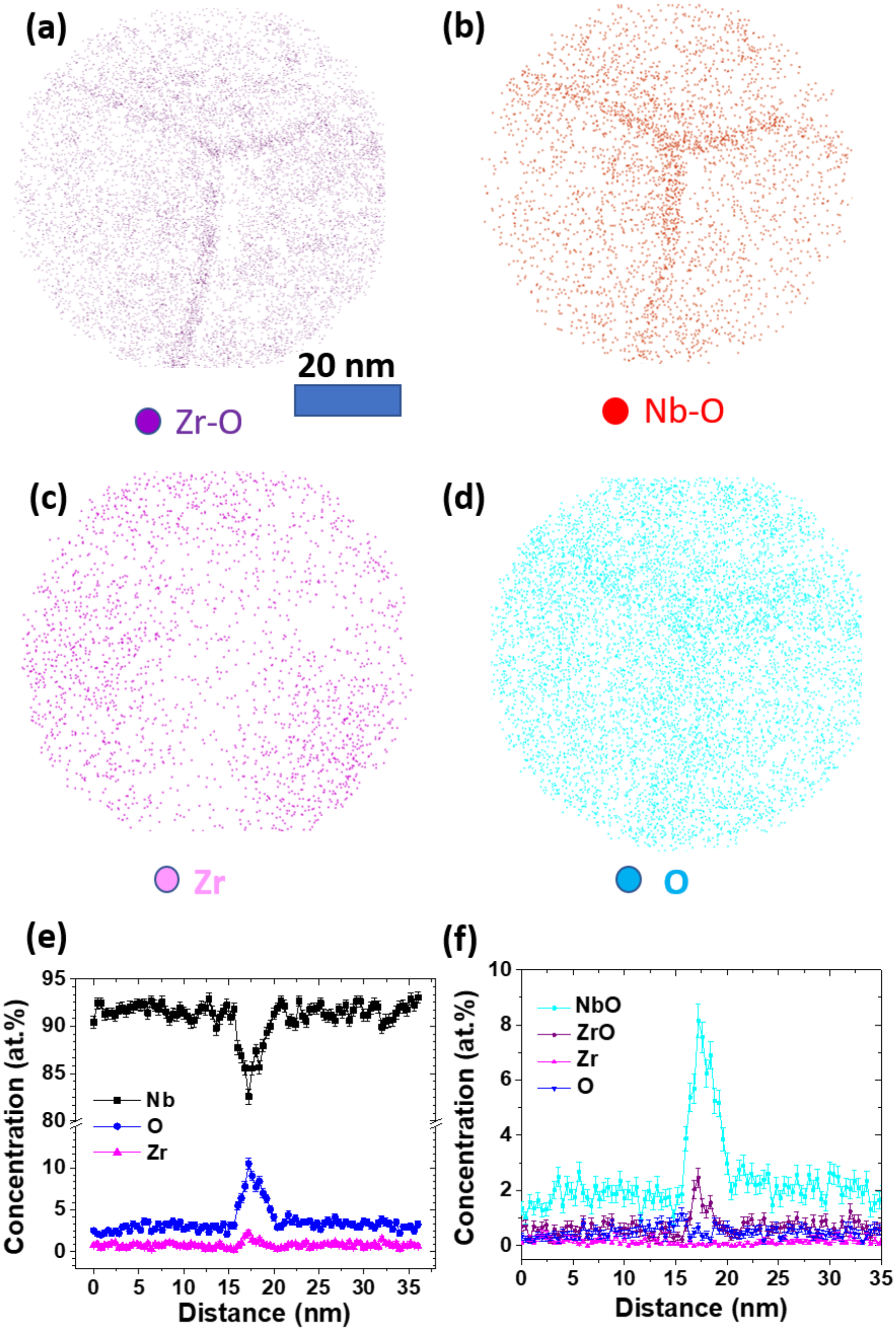

**Figure 2**. 2-D sections through 3D APT reconstructions of a nanotip of the unreacted Nb alloy region with grain boundaries illustrate the distributions of (a) Zr-O and (b) Nb-O molecular ions, and (c) atomic Zr, and (d) atomic O ions in the APT reconstruction. Plots for (e) atomic concentration profiles of Nb, O, Zr and (f) ionic concentration profiles of Nb-O and Zr-O molecular ions, with profiles for atomic Zr, and atomic O across a GB are provided. They reveal that Zr-O and Nb-O dimers are segregated at Nb GBs.

### 3.3 Transmission electron microscopy and atom-probe tomographic analyses of $Nb_3Sn$/Nb interfaces

A bright-field (BF)-STEM image, **Fig. 3(a),** exhibits an inter-reaction interface between $Nb_3Sn$ and Nb for a sample heat-treated at 625 °C. The Nb grain is oriented close to a [110] zone axis and thus diffracts strongly. Therefore, the contrast of Nb in BF-STEM Nb is darker than $Nb_3Sn$. This $Nb_3Sn$/Nb interface exhibits an undulating morphology rather than an atomically flat sharp interface.

Details of the distributions of atomic and molecular ions in the region at a $Nb_3Sn$/Nb heterophase interface are displayed in cross-sectional slices through an APT reconstruction containing a $Nb_3Sn$/Nb interface, **Figs. 3 (b-d)**. Preliminary analyses of the concentrations of O, Cu, and Sn, and their diffusion along an Nb/$Nb_3Sn$ interface were reported in ref. [19]: the current study focuses on the nucleation of Zr oxide nanoprecipitates at a Nb/$Nb_3Sn$ hetero-phase interface. Herein, we observe a distinct difference between Zr-O distributions in Nb and in $Nb_3Sn$. In Nb, Zr and O remain more or less uniformly distributed in solid solutions with potentially individual Nb-O or Zr-O dimers. On the other hand, Zr-O is aggregated in more discrete clusters in $Nb_3Sn$, in other words, they begin to form precipitates. Zr-O distribution in the 3D APT reconstruction, **Fig. 3**, reveals that the concentration of Zr-O dimers is higher at $Nb_3Sn$/Nb interfaces and $Nb_3Sn$ GBs compared to that in the bulk of $Nb_3Sn$ grains, indicating that Zr oxide precipitates actively nucleate at the $Nb_3Sn$/Nb interfaces and in $Nb_3Sn$ GBs. **Fig. 3(c,d)** displays the distribution of Zr-O and Nb-O dimers across the $Nb_3Sn$/Nb interface. In the 3-D APT reconstructions of the location of Zr-O and Nb-O dimers, Zr-O is enriched in the same regions as Nb-O dimers in Nb and $Nb_3Sn$, indicative of the incorporation of Nb in Zr oxide precipitates, which is attributed to the existence of the excess O and high affinity of O to Nb, **Table S1.** We also note a high Sn concentration at Nb GBs close to the $Nb_3Sn$/Nb inter-reaction interface, indicating that Sn diffuses preferentially into the Nb GBs at the Nb/$Nb_3Sn$ interface, **Fig. 3(b)**. Additionally, the protrusion of the $Nb_3Sn$ layer into Nb GBs suggests that Nb GBs are preferential sites for the growth of $Nb_3Sn$.

The atomic concentration profiles across the $Nb_3Sn$/Nb interface was presented in ref. [19] using the proximity histogram methodology [26] for calculating concentration profiles across interfaces with arbitrary and undulating geometries. It demonstrates that the oxygen concentration is greater in Nb (5.5 ±0.2 at.% O) than in $Nb_3Sn$ (2.1 ±0.1 at.% O), which is also indicated in the ionic concentration profile across the $Nb_3Sn$/Nb interface, **Fig. S2.** This leads to an O concentration gradient from Nb to $Nb_3Sn$ as exhibited in the concentration profile across the $Nb_3Sn$/Nb interface. The high concentration of oxygen observed at the $Nb_3Sn$/Nb heterophase interface in **Fig. S2** and **ref. [19]**, is attributed to two factors: (i) an active nucleation of Zr oxide nanoprecipitates at this interface; and (ii) the accumulation of oxygen atoms at the moving inter-reaction heterophase interface due to the small solubility of oxygen in $Nb_3Sn$. Details of the kinetics of interfacial reactions are explained in the discussion section. We note that Cu is segregated at the interface due to the fast diffusion of Cu along the $Nb_3Sn$/Nb heterophase interface.

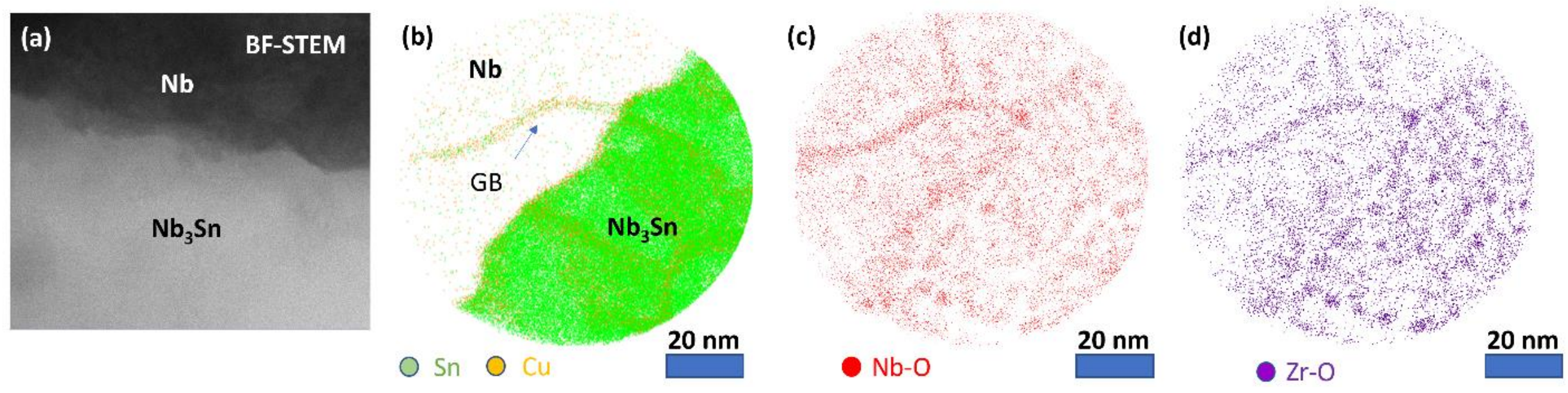

**Figure 3.** (a) BF-STEM image of the $Nb_3Sn$/Nb interface in an internally oxidized $Nb_3Sn$ wire heat treated at 625 °C exhibits an undulating irregular morphology of the interface. (b-d) Cross-sectional slices through a 3D APT reconstruction capturing a $Nb_3Sn$/Nb inter-reaction interface with adjacent Nb and $Nb_3Sn$ grains in a nanotip of the internally oxidized $Nb_3Sn$ wire reacted at 625°C: (b) distribution of Sn and Cu atoms near the $Nb_3Sn$/Nb heterophase interface: Sn and Cu

diffusion along the GBs of Nb are observed. Distributions of (c) Nb-O and (d) Zr-O molecular ions at a Nb/$Nb_3Sn$ heterophase interface demonstrate that Nb-O and Zr-O molecular ions detected in APT are originating from locations at GBs in Nb, the Nb/$Nb_3Sn$ interface, GBs in $Nb_3Sn$, and discrete (Zr, Nb)-oxide precipitates in $Nb_3Sn$.

## 3.4 Transmission electron microscopy and atom-probe tomography analyses of Zr oxide precipitates in $Nb_3Sn$ layers

### *3.4.1 Scanning transmission electron microscopy analyses of Zr oxide precipitates in $Nb_3Sn$*

We now describe the evolution of Zr oxide nanoprecipitates during the growth of $Nb_3Sn$ layers based on the results from S/TEM and APT analyses. BF-STEM images in **Fig. 4(a,b)** display Zr oxide precipitates in $Nb_3Sn$ filaments at different distances from Nb/$Nb_3Sn$ heterophase interfaces for the heat treatment temperatures 625 °C and 700 °C, respectively. The average diameters of Zr oxide nanoprecipitates are measured employing BF-STEM images using ImageJ software [27]. For both reaction temperatures, moderate coarsening of Zr oxide nanoprecipitates with distance from Nb/$Nb_3Sn$ heterophase interfaces is observed. Coarsening is more significant at 700 °C due to the greater diffusivities of Zr and O in the $Nb_3Sn$ matrix at 700 °C. At 1 μm distance from the Nb/$Nb_3Sn$ heterophase interface, the average diameter of the Zr oxide nanoprecipitates is 3.1 ±0.5 nm for both the 625 °C and 700 °C samples, with the stated error representing the standard deviation of the distribution. Further away from the interface at 5 μm distance, the average precipitate diameter increases to 6.8 ±1.8 nm at 625 °C and 9.1 ±3.9 nm at 700 °C, respectively. There is a larger standard deviation of the nanoprecipitate diameter distribution at 700 °C, due to the presence of some larger nanoprecipitates (~15 nm). Nevertheless, the majority of the nanoprecipitates have diameters <10 nm, which is consistent with the observations from APT analyses, **Fig. S3**.

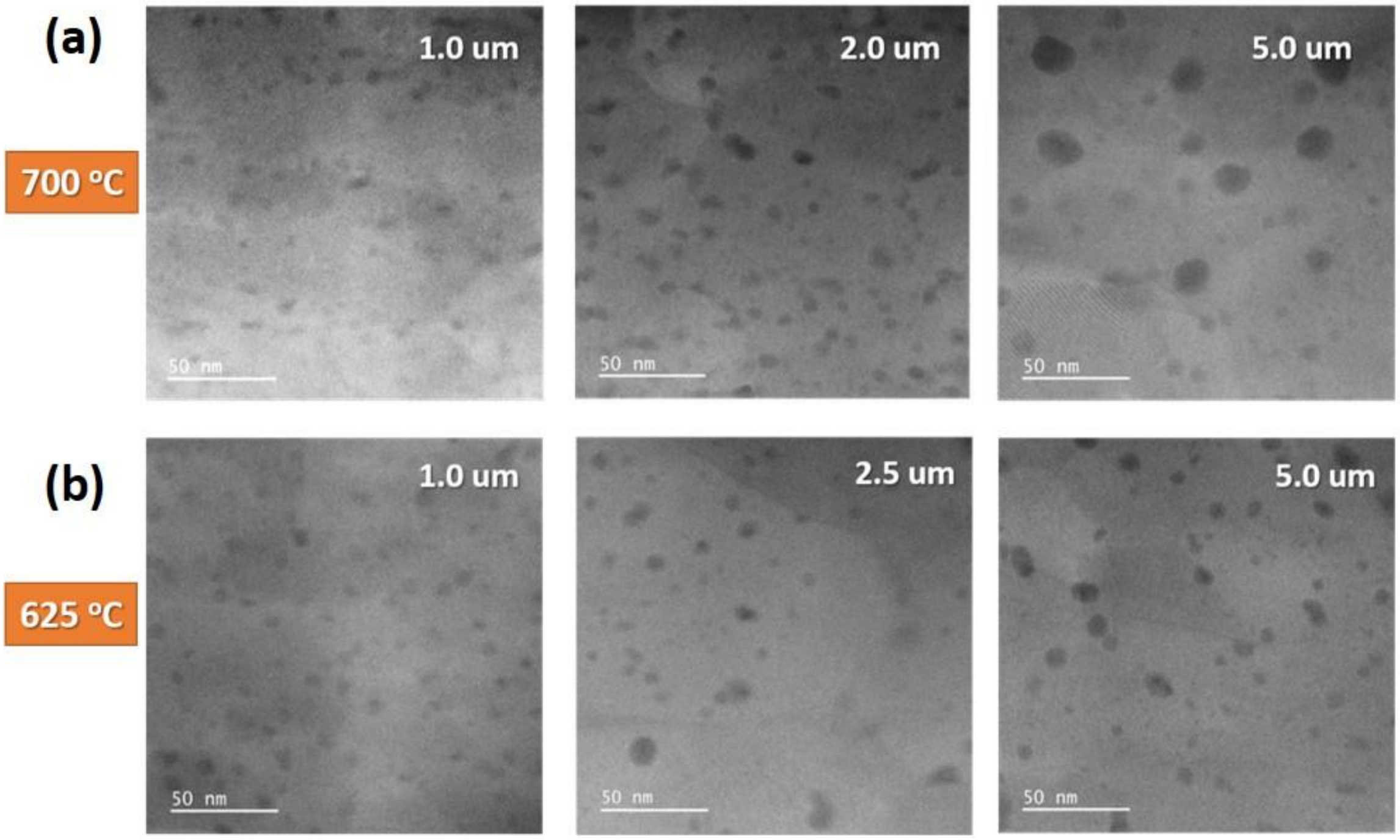

**Figure 4**. BF-STEM images of Zr oxide precipitates at different distances (0.2-5.0 μm) from the Nb/$Nb_3Sn$ inter-reaction interface in the internally oxidized $Nb_3Sn$ samples heat-treated at (a) 700 °C for 120 h and (b) 625 °C for 740 h. The Zr oxide precipitates coarsen moderately with increasing distance from the Nb/$Nb_3Sn$ inter-reaction interface. The coarsening is more significant at 700 °C compared to 625 °C.

High-resolution STEM imaging is employed to analyze the structure of selected Zr oxide precipitates. According to the Zr-O phase diagram, the equilibrium phase of Zr oxide at 625-700 °C has a monoclinic crystal structure, m-$ZrO_2$. Other alloying elements can, however, transform and stabilize the m-$ZrO_2$ into a cubic structure [28]. Additionally, the A15 cubic structure of the $Nb_3Sn$ matrix, in which the Zr oxide nanoprecipitates are embedded, can influence the structure of the Zr oxide nanoprecipitates. High-resolution bright-field STEM images, **Fig. 5,** display the atomic structure of Zr oxide nanoprecipitates imaged along the $Nb_3Sn$ [111] zone axis. The crystal structure of the Zr oxide nanoprecipitates is close to cubic, c-$ZrO_{2-x.}$ ($0 \leq x \leq 2$), with a lattice constant of 5.4 Å. This value of the lattice constant is ~5% larger than the archival literature value of cubic $ZrO_2$, 5.15 Å (JCPDS: 27-0997), which can be attributed to the presence of Sn and Nb in the Zr oxide precipitates. Fast Fourier Transformations (FFTs) of the BF-STEM images and filtered BF-STEM images, **Fig. 5(a,b)**, reveal the orientation relationship for the c-$ZrO_{2-x}$/$Nb_3Sn$ heterophase interface: c-$ZrO_{2-x}$(110) // $Nb_3Sn$ (110) and c-$ZrO_{2-x}$($1\bar{1}0$) // $Nb_3Sn$ (112). We note that the Zr oxide nanoprecipitates ($ZrO_{2-x}$ with $0 \leq x \leq 2$) can exist as non-equilibrium phases due to the non-equilibrium state of the current system [29, 30]. Additionally, atomic resolution images of the Zr oxide nanoprecipitates reveal lattice distortions and the presence of dislocations, which can play additional roles in the formation of non-equilibrium c-$ZrO_x$ precipitates ($0 \leq x \leq 2$) embedded in $Nb_3Sn$.

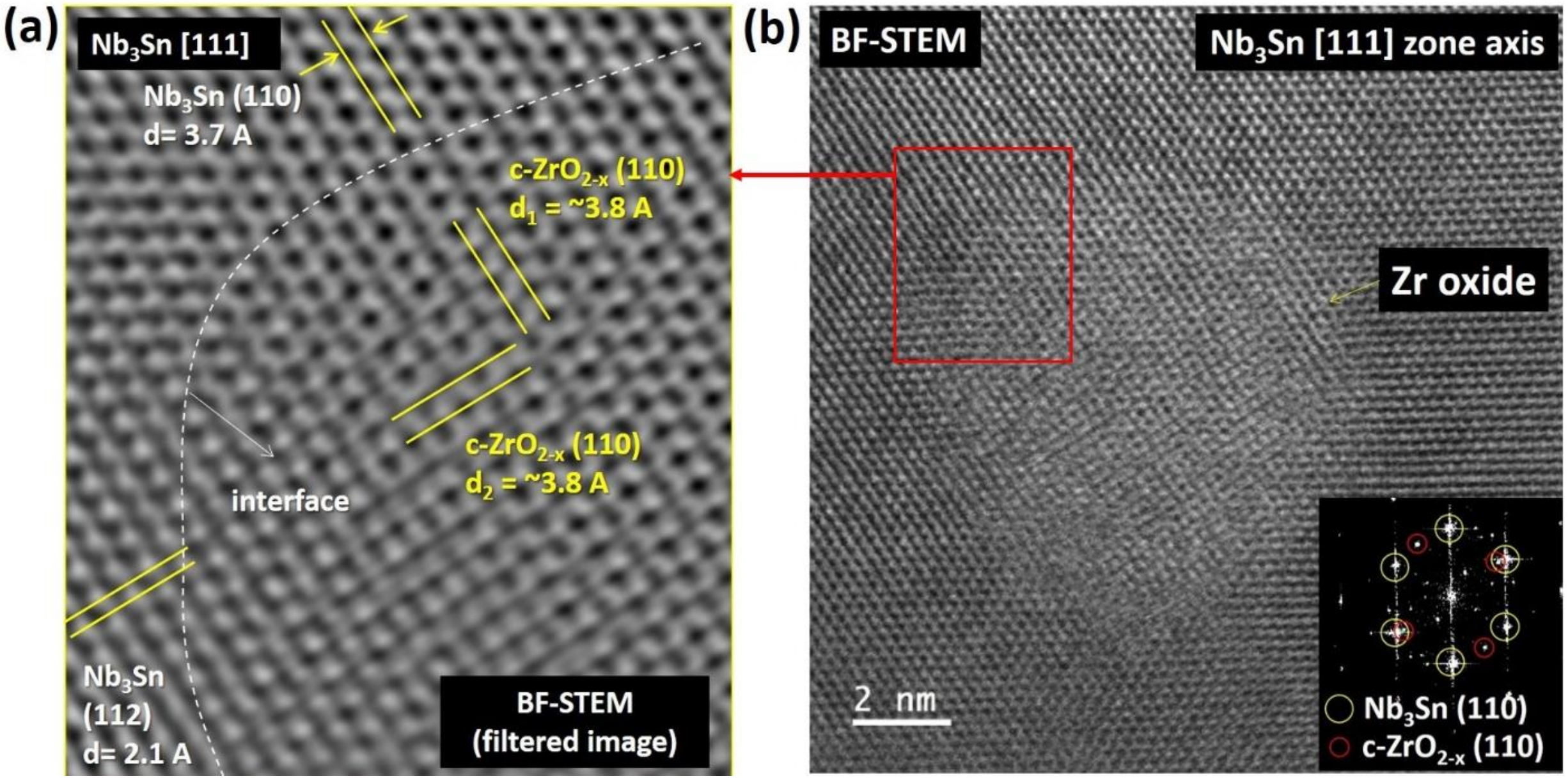

**Figure 5**. (a) High-resolution BF-STEM and Fast Fourier Transformed (FFT) images of a Zr oxide precipitate in a $Nb_3Sn$ grain imaged along a [111] zone axis reveal the atomic structure of the Zr oxide precipitate, which is close to cubic, c-$ZrO_{2-x.}$ ($0 \leq x \leq 2$), with a lattice constant of 5.4 Å. (b) Filtered BF-STEM image of the Zr oxide precipitate and $Nb_3Sn$ matrix shows that the Zr oxide precipitate forms a coherent interface with the $Nb_3Sn$ matrix with orientation relationship c-$ZrO_{2-x}$(110) // $Nb_3Sn$ (110) and c-$ZrO_{2-x}$($1\bar{1}0$) // $Nb_3Sn$ (112).

*3.4.2 Atom probe tomographic analyses of Zr oxide nanoprecipitates in $Nb_3Sn$*

APT is utilized to provide detailed analyses of the number densities, nanoprecipitate diameters, and chemical compositions of Zr oxide precipitates. The cluster analysis module based on a maximum-separation algorithm in the IVAS software program (Cameca, Madison, WI) is employed to extract the information as a function of distance from Nb/$Nb_3Sn$ heterophase interfaces [31, 32].

**Fig. 6** presents the number densities ($N_v$) and average diameters of Zr oxide nanoprecipitates at 625 °C, which illustrates the nucleation, growth, and coarsening behavior of nanoprecipitates as a function of distance from the $Nb_3Sn$/Nb inter-reaction interface. For the number density ($N_v$) and diameter of Zr oxide precipitates at a distance of 4.5 µm, **Fig. 6,** denoted by ”*”, TEM was employed as the Zr oxide precipitates are large enough (> ~5 nm) to be resolved in TEM images. The thickness of the TEM foil in $Nb_3Sn$ regions is taken to be 50 nm to estimate the number density of Zr oxide precipitates. This APT analysis discovers that active nucleation and growth of Zr oxide nanoprecipitates occur within a distance of ~200 nm from Nb/$Nb_3Sn$ heterophase interfaces. At this stage, the number density of Zr oxide precipitates achieves a value as high as $1.5\times10^{24}$ $m^{-3}$, **Fig. 6**. At a distance of ~1 µm from Nb/$Nb_3Sn$ heterophase interfaces, the $N_V$ of Zr oxide precipitates decreases to $7.6\times10^{23}$ $m^{-3}$, which is an indicator of coarsening of Zr oxide precipitates. Subsequently, the Zr oxide nanoprecipitates continue coarsening and number density ($N_V$) decreases further to $3.2\times10^{22}$ $m^{-3}$ at 4.5 µm from the Nb/$Nb_3Sn$ heterophase interfaces. Meanwhile, the average diameter of Zr oxide nanoprecipitates increases from 1.3±0.2 nm near the Nb/$Nb_3Sn$ interfaces to 7.1 ±0.1 nm at a distance of 4.5 µm from the interface.

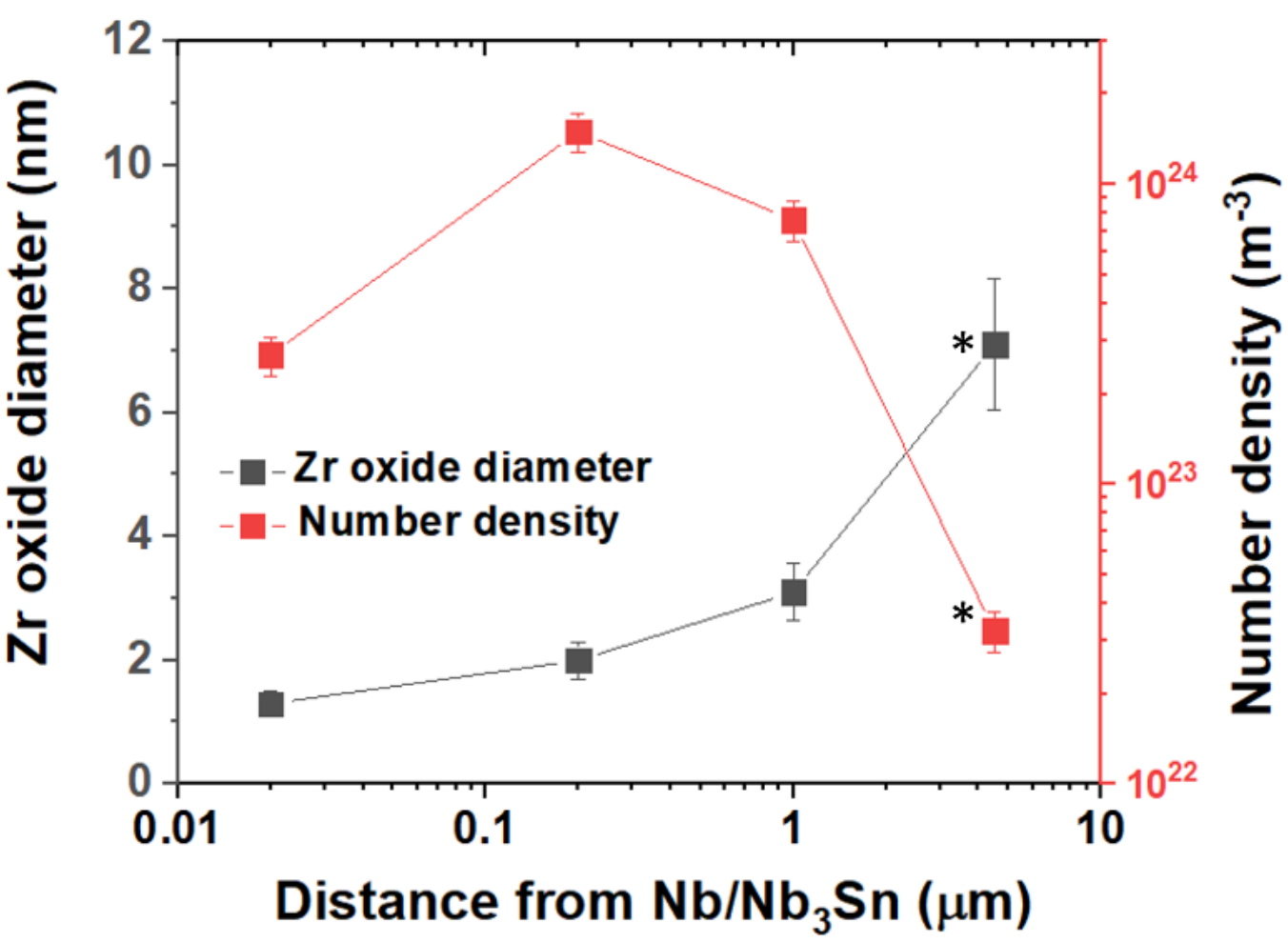

**Figure 6**. Average diameters and number densities ($N_v$) of Zr oxide precipitates in $Nb_3Sn$ heat-treated at 625 °C for 740 h, as a function of distance from the Nb/$Nb_3Sn$ interface, determined by APT, except the data point at a distance of 4.5 μm from the Nb/$Nb_3Sn$ interface which was estimated by TEM and is denoted by "*".

Next, we analyze the chemical composition of Zr oxide nanoprecipitates in a sample heat-treated at 625 °C utilizing APT. Specifically, Zr oxide nanoprecipitates at different distances from Nb/$Nb_3Sn$ heterophase interfaces up to ~1 μm were selected for systematic analyses, as indicated in **Fig. 1(b)**. To examine the compositional evolution of Zr oxide nanoprecipitates during nucleation and growth, atomic concentration profiles across Zr oxide/$Nb_3Sn$ heterophase interfaces are calculated using the proximity histogram subroutine in IVAS and displayed in **Fig. 7(a-c)** [26]. We note that Zr oxide nanoprecipitates contain Nb and Sn atoms at the nucleation stage, **Figs. 7**, and as the growth process progresses, they evolve toward the equilibrium composition, $ZrO_2$. Concentration profiles of molecular ions across Zr oxide/$Nb_3Sn$ interfaces, **Fig. 7(d-f)**, also show an increase of Zr-O molecular ions and a decrease of Nb-O molecular ions in Zr oxide precipitates with increasing distance from the Nb/$Nb_3Sn$ interface. Additionally, a small amount of Cu incorporation (~1.5 at.%) in Zr oxide precipitates is observed, **Fig. S3**. The chemical evolution of Zr oxide precipitates is influenced by several thermodynamic and kinetic factors, such as chemical interactions among solute species, which are discussed in the **Supplementary Information, Table S1** and **Fig. S4**.

A representative 3D reconstruction of a nanotip containing $Nb_3Sn$ grains with Zr oxide nanoprecipitates is presented in **Fig. 7(g)**, which is at a distance of ~1 μm from $Nb/Nb_3Sn$ heterophase interfaces for the 625 °C sample. Iso-concentration surfaces of 2.5 at.% of Zr are selected to identify Zr oxide nanoprecipitates in the $Nb_3Sn$ matrix. The 3D APT reconstructed image reveals a distribution of Zr oxide nanoprecipitates in $Nb_3Sn$ grains, including the presence of large precipitates at $Nb_3Sn$ GBs. It is noteworthy that Zr oxide nanoprecipitates at $Nb_3Sn$ GBs are generally larger than those inside $Nb_3Sn$ grains because the diffusivities of Zr and O along GBs are greater than those in the $Nb_3Sn$ grain interior. The number density ($N_v$) of Zr oxide nanoprecipitates in the APT nanotip, ~1 μm away from the $Nb_3Sn/Nb$ interface, is estimated to be $7.6 \times 10^{23}$ $m^{-3}$.

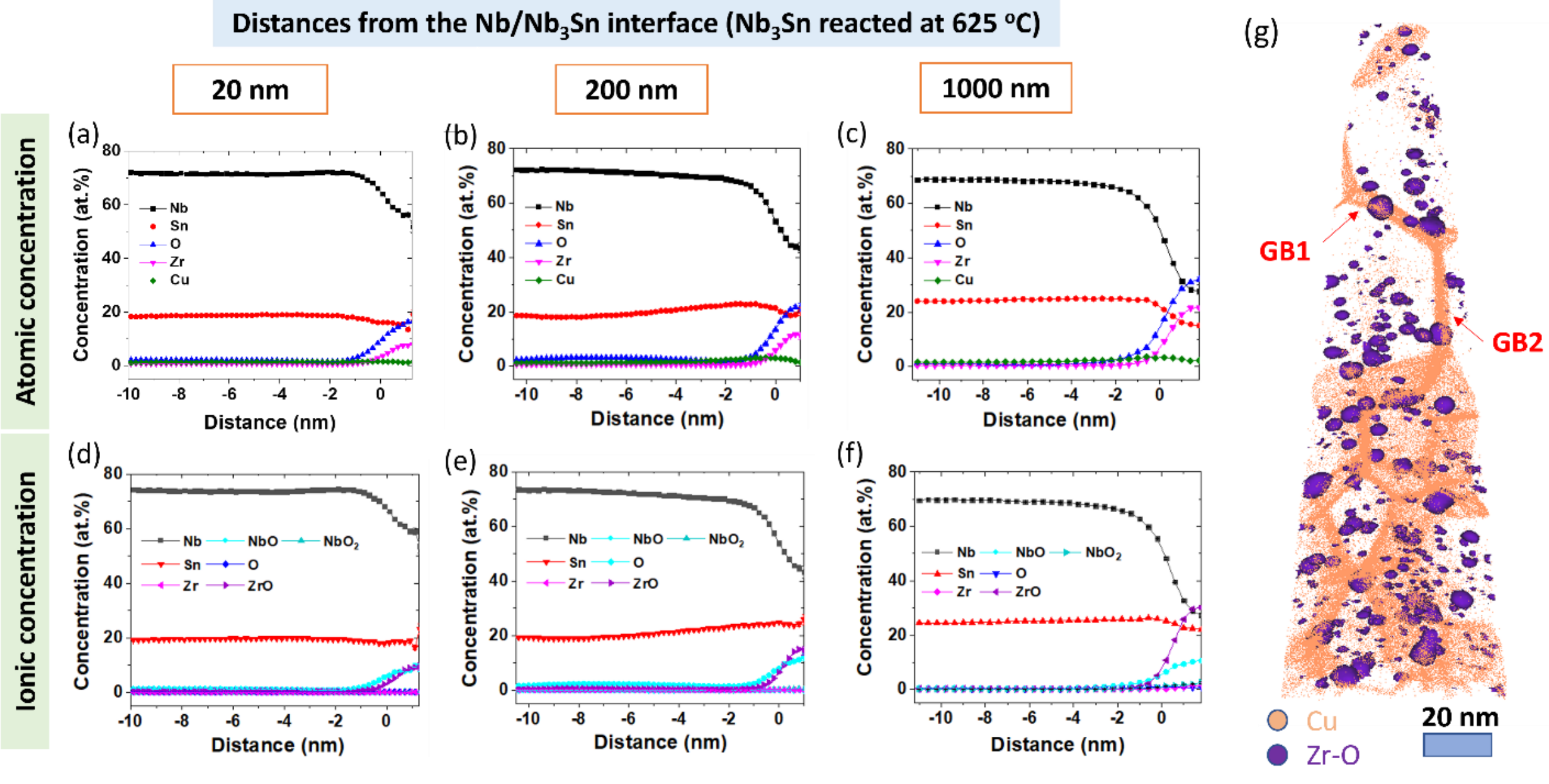

**Figure 7.** (a-c) Atomic and (d-f) ionic concentration profiles across the Zr oxide precipitate/$Nb_3Sn$ interface at different distances from the $Nb/Nb_3Sn$ interface of the $Nb_3Sn$ wire reacted at 625 °C

for 740 h, determined with the proximity histogram methodology in IVAS software. In combination, the concentration profiles illustrate the compositional evolution of Zr oxide precipitates during the growth of the $Nb_3Sn$ layer at 625 °C: they evolve toward the equilibrium composition, $ZrO_2$. (g) 3-D APT reconstruction of the $Nb_3Sn$ nanotip with Zr oxide precipitates at ~1 μm away from the $Nb_3Sn$/Nb interface is provided. Cu is segregated at $Nb_3Sn$ GBs and finely-dispersed Zr oxide precipitates with an average diameter (3.1 ±0.5 nm) and with high number density ($7.1 \times 10^{23}$ $m^{-3}$) are seen: iso-concentration surface of Zr-O molecular ion (2.5 %) is superimposed to delineate the Zr oxide precipitates.

Additionally, we performed APT analyses of the $Nb_3Sn$ sample heat treated at 700 °C for 120 h at a distance of ~5 μm away from the Nb/$Nb_3Sn$ heterophase interface, **Fig. S3**, to study the effect of a higher heat-treatment temperature. The chemical composition of Zr oxide nanoprecipitates is estimated to be 25.7 ±2.5 at.% Zr, 36.5 ±2.8 at.% O, 21.1 ±2.3 at.% Nb, 13.2 ±1.9 at.% Sn, 1.5 ±0.1 at.% Cu. Compared to the Zr oxide nanoprecipitates in the sample heat-treated at 625 °C, the composition of Zr oxide nanoprecipitates at 700 °C is closer to stoichiometric $ZrO_2$. We note that we evaluated a possible local magnification effect at the Zr oxide nanoprecipitates/$Nb_3Sn$ interfaces and concluded it is insignificant, which is summarized in **Supplementary information**, **Fig. S5** [33, 34].

### 3.5 Theoretical analyses of the nucleation of $ZrO_2$ in $Nb_3Sn$ using classical nucleation theory and first-principles calculations

We now employ classical nucleation theory (CNT) to discuss the nucleation mechanism of Zr oxides in $Nb_3Sn$ and to estimate the nucleation energy barrier (net reversible work, $W(R)$, to form a critical nucleus) and critical radius ($r^*$) of a nucleus. Here, CNT and first-principles calculations are performed assuming homogeneous nucleation of cubic $ZrO_2$ in $Nb_3Sn$ to understand the general nucleation behavior of Zr oxide precipitates in $Nb_3Sn$ based on the HR-STEM analysis, **Fig. 5**. For homogeneous nucleation of a spherical precipitate, the net reversible work for the nucleation of a spherical precipitate, $W(r)$ with a radius $r^*$ is determined by three factors: (i) chemical Helmholtz free energy change during nucleation ($\Delta F_{chem}$); (ii) elastic strain energy change ($\Delta E_{elastic}$); and (iii) interfacial free energy at the precipitate/matrix interface ($\sigma$) [35, 36]:

$$W(r) = \frac{4\pi}{3} r^3 \Delta F_v + 4\pi r^2 \sigma \quad (1)$$

The critical radius of the nucleus ($r^*$):

$$r^* = -2\sigma / \Delta F_v \quad (2)$$

where

$$\Delta F_v = \Delta F_{chem} + \Delta E_{elastic} \quad (3)$$

The free energy change for nucleation ($\Delta F_v$) consists of two terms, bulk formation energy ($\Delta F_{chem}$) and strain energy ($\Delta E_{elastic}$). The details of the first-principles calculations of each term are based on TEM analyses of the crystal structure of $ZrO_2$ and orientation relationships at $Nb_3Sn/ZrO_2$, are explained in **Supplementary Section** and **Fig. S6**.

We finally estimate $W(r^*)$ and $r^*$ for homogeneous nucleation of cubic $ZrO_2$ based on the calculated bulk formation Helmholtz free energy change ($\Delta F_{chem}$) for the nucleation of $ZrO_2$ from the $Nb_3Sn$-Zr-O mixture, strain energy ($\Delta E_{elastic}$), and interfacial energy ($\sigma$) at a $ZrO_2/Nb_3Sn$ interface, see details in supplementary information S3. The calculated $r^*$ for nucleation of cubic $ZrO_2$ is 1.2 nm at 625 °C and 1.4 nm at 700 °C, respectively. $W(r^*)$ for the formation of a spherical precipitate is estimated to be 3.71 $\times 10^{-21}$ J at 625 °C and 4.43 $\times 10^{-21}$ J at 700 °C, respectively. It is worth noting that both the energy barriers ($W(r^*)$) and critical radius ($r^*$) are larger at higher temperatures due to an increased contribution of vibrational entropy, **Fig. S6**. The calculated values from the first-principles calculations are summarized in **Table 1**.

**Table 1**. List of calculated values for the classical nucleation model using first-principles calculations

| | Temperatures | | |
|---|---|---|---|
| | 0 K | 625 °C | 700 °C |
| Critical radius ($r^*$) | - | 1.2 nm | 1.4 nm |
| Nucleation barrier (net reversible work, $W(r)$) | - | $3.71 \times 10^{-21}$ J | $4.43 \times 10^{-21}$ J |
| Interfacial energy at (110) $ZrO_2$/(110) $Nb_3Sn$ ($\sigma$) | 325 mJ/m$^2$ | - | - |
| Strain energy at (110) $ZrO_2$/(110) $Nb_3Sn$ ($E_{strain}$) | 0.194 eV/atom | - | - |
| Zr substitutional energy to Nb site | 0.124 eV/atom | - | - |
| Zr substitutional energy to Sn site | 0.475 eV/atom | - | - |

## 4 Discussion

In this study, we investigate the evolution of Zr oxide precipitates in internally oxidized $Nb_3Sn$ wires. To achieve this, we employ APT and TEM, and they are complemented by first-principles calculations. This section discusses the details of the nucleation and growth of Zr oxide precipitates in $Nb_3Sn$.

### 4.1 Formation of Zr-O clusters in an unreacted Nb alloy

In **Fig. 2**, we observe significant clustering of Zr-O in the unreacted Nb alloy: this clustering is a result of the strong affinity of Zr for O. The excess Zr and O solutes above the solubility limit in Nb segregate at Nb GBs, which result in high concentrations of Zr-O and Nb-O clusters at Nb GBs. The binary Nb-Zr system has a miscibility gap in the solid solution phase and Zr atoms are segregated at GBs, which reduces the free energy of the Nb-Zr solid solution [37, 38]. According to the Nb-Zr-O ternary phase diagram, the equilibrium state for a given composition of Nb-Zr-O (1.0 ±0.1 at.% Zr, 5.5 ±0.2 at% O, Nb = balance) at 625 °C consists of a mixture of three phases: Nb, NbO, $ZrO_2$ [39]. In our Nb-Zr-Ta alloy, 4 at.% of Ta is added to enhance the upper critical field ($B_{c2}$) and Ta dissolves in the Nb matrix [12, 40]. We neglect, for simplicity, the presence of Ta when predicting the equilibrium phases. We haven't observed evidence for crystalline $ZrO_2$ or NbO precipitates in our Nb-Zr-Ta-O alloy in the TEM and APT analyses and this is rationalized by the small diffusivity of Zr in Nb [39].

We estimate the root-mean-square diffusion distance and annealing time required for Zr diffusion to initiate the nucleation of $ZrO_2$ precipitates in Nb. Employing our DFT calculations and classical nucleation theory (CNT), **Table 1**, the number of Zr atoms involved in nucleation is calculated using the formula: $N_a \times 4\pi r^{*3}/(3V_m) = 200.4$ (where $N_a$ is Avogadro's constant, $6.02 \times 10^{23}$/mol, and $V_m$ is the molar volume of $ZrO_2$, 21.7 $cm^3$/mol, based on its molar mass and mass density, and $r^*$ is critical radius of Zr oxide precipitates, 1.2 nm). Considering that each unit cell of bcc Nb contains two Nb atoms and an average of 0.02 Zr atoms (as the concentration of Zr in the Nb-Zr solid solution is 1%), all the Zr atoms in the surrounding 10,022 unit cells of bcc Nb needs to diffuse to form the nucleus of 1.2 nm $ZrO_2$. With a lattice constant of 0.33 nm for the Nb-1%Zr solid solution, we can calculate the required diffusion distance for Zr solute atoms, 4.5 nm. The diffusion of Zr in Nb is extremely slow, where $D_{Zr}$ in Nb at 625 °C is estimated to be $1.3 \times 10^{-24}$ $m^2$/s [41, 42], and the estimated diffusion length of Zr in Nb for 740 h at 625 °C ($\sqrt{4D_{Zr}t}$) is 3.8 nm, which is ~15% less than the required diffusion length for the nucleation of $ZrO_2$ in Nb. Therefore, Zr solute atoms may remain bound to O as a dimer in solid solution rather than forming distinct $ZrO_2$ precipitates in Nb.

### 4.2 Nucleation of Zr oxide precipitates at the Nb/$Nb_3Sn$ interface and in $Nb_3Sn$

In **Fig. 3(c)**, we observe active nucleation of Zr oxide precipitates at a Nb/$Nb_3Sn$ interface. This is attributed to the significant reduction in solubilities of Zr and O in $Nb_3Sn$ compared to Nb. While there isn't a precise phase diagram for the Nb-Sn-Zr-O quaternary system, the solubility of O in $Nb_3Sn$ has been estimated by calculating the formation energy of oxygen interstitials in $Nb_3Sn$ using first-principles calculations. This result indicates that the solubility of oxygen in $Nb_3Sn$ is ~5 orders of magnitude smaller than in Nb at ~1000 K [43], due to the much larger formation energy for O interstitials in $Nb_3Sn$ than in Nb. As a result, the concentrations of O in the $Nb_3Sn$ matrix are substantially smaller than in Nb across the interface, **Fig. 3(d)** and ref. [19].

Also, we observe a significant amount of heterogeneous nucleation of Zr oxides at Nb/$Nb_3Sn$ interfaces and at $Nb_3Sn$ GBs. The nucleation barrier ($W(r)$) and critical radius ($r^*$) values calculated using first-principles calculations, **Table. 1**, assume homogeneous nucleation, and therefore, the actual nucleation barrier ($W(r)$) and critical radius ($r^*$) of Zr oxide precipitates at the Nb/$Nb_3Sn$ interface and grain boundaries in $Nb_3Sn$ would be smaller than the calculated values in **Table. 1**. The presence of Nb/$Nb_3Sn$ heterophase interfaces and $Nb_3Sn$ GBs aid the heterogeneous nucleation of Zr oxide precipitates at these interfaces and significantly promote the nucleation of Zr oxides in $Nb_3Sn$.

We emphasize that $W(r^*)$ and $r^*$ for nucleation of Zr oxide precipitates in $Nb_3Sn$ increases as the temperature increases. This is attributed to the larger contribution of phonon vibration entropies at higher temperatures. Consequently, $W(r^*)$ for the nucleation of Zr oxide precipitates is small at 625 °C compared to 700 °C. At lower temperatures, the diffusivities of Zr and O are small, which limits the nucleation of $ZrO_2$ precipitates. Therefore, it is necessary to increase the temperature above a threshold value to ensure that the diffusivities of Zr and O are large enough to form nuclei during a heat treatment.

### 4.3 Effects of Zr oxide precipitates on the microstructure of $Nb_3Sn$ layers

The high number density ($N_v$) of Zr oxide nanoprecipitates in $Nb_3Sn$ has a significant impact on the microstructural evolution of $Nb_3Sn$ layers in internally oxidized $Nb_3Sn$ wires. There is a noticeable grain refinement observed at both 625 °C and 700 °C, relative to $Nb_3Sn$ wires without oxide nanoprecipitates, resulting in a high-volume density of GBs for pinning vortices to enhance $J_c$. The increase of the average grain diameter in $Nb_3Sn$ after heat treatments is relatively small from 70±12 nm to 100±18 nm as the heat treatment temperature increases from 625 °C to 700 °C. This indicates that the presence of Zr oxide nanoprecipitates effectively impedes grain growth of $Nb_3Sn$ [23]. By considering the average radius ($r$) and volume fraction ($f$) of Zr oxide precipitates, a maximum radius of grains ($R$), $R = 4r/(3f)$ is calculated to be ~60.5 nm, and in turn, a maximum diameter of grains is ~121 nm, assuming $2r = 9.1$ nm and $f = 0.10$ for the 700°C/120 h sample based on the BF-STEM image (**Fig. 4**) [17, 23]. This value is close to the grain diameter of $Nb_3Sn$ observed at 700 °C for 120 h.

Another interesting aspect of the $Nb_3Sn$ formation is the preferential growth of $Nb_3Sn$ along Nb GBs at the Nb/$Nb_3Sn$ heterophase interface, **Figs. 3.** Prior studies have highlighted the importance of the microstructure of Nb grains, such as the grain diameter of Nb, on the microstructure of the $Nb_3Sn$ layers [13, 44]. This observation suggests that the microstructure of Nb plays a role in the nucleation and growth of $Nb_3Sn$ grains with Nb GBs providing short-circuit paths for Sn diffusion. During the wire fabrication process, Nb grains undergo severe deformation due to repeated cold drawing, which leads to a small Nb grain diameter.

### 4.4 Insights for improving the superconducting properties of internally oxidized $Nb_3Sn$ wires with a high density of magnetic flux pinning centers

Our studies on the nucleation and growth of Zr oxide precipitates offer valuable insights for engineering the diameter and number density ($N_v$) of Zr oxide nanoprecipitates in the internally oxidized $Nb_3Sn$ superconducting wires to improve their flux pinning efficiency. For instance, we showed that Zr-O segregation occurs at GBs in Nb, serving as a precursor to the precipitation of Zr oxide nanoprecipitates and helping the nucleation of Zr oxide nanoprecipitates at Nb/$Nb_3Sn$ heterophase interfaces. We also find that the oxygen concentrations in Nb and $Nb_3Sn$ play important roles in the incorporation of Nb into Zr oxides and their temporal evolution. By controlling the oxygen concentration to obtain the stoichiometry of $ZrO_2$, we may be able to reduce the concentration of Nb in Zr oxide nanoprecipitates. This can lead to a reduction in the diameter and coarsening rate of Zr oxide nanoprecipitates. By controlling these factors, we can potentially enhance the pinning effects and overall superconducting performance of $Nb_3Sn$ wires. Furthermore, the systematic chemical analyses of nanoprecipitates, the number density of Zr oxide nanoprecipitates, and compositions of the $Nb_3Sn$ layers can provide important information for the possible Ginzburg-Landau simulation of $J_c$ in the $Nb_3Sn$ layers with different arrangements of pinning centers [45, 46]. Also, the calculated values in DFT and CNT of the current studies provide important parameters to demonstrate Zr oxide precipitates precipitation in $Nb_3Sn$ layers as a function of time and distance using phase-field modeling.

## 5. Conclusions

Our research describes the kinetic pathways of the nucleation and growth of Zr oxide nanoprecipitates and the microstructural evolution of $Nb_3Sn$ layers in internally oxidized $Nb_3Sn$ wires. To achieve this, we utilize APT, TEM, and first-principles calculations at the atomic scale. The presence of Zr oxide nanoprecipitates in internally oxidized $Nb_3Sn$ wires plays critical roles in the microstructural temporal evolution of $Nb_3Sn$ grains, acting as Zener pinning centers in the process of $Nb_3Sn$ grain growth and directly providing pinning centers for magnetic vortices. This research elucidates a series of important aspects for the evolution of Zr oxide nanoprecipitates and the control of the microstructure of internally oxidized $Nb_3Sn$ wires.

- The APT analyses of the unreacted Nb alloy volume reveal the distribution and clustering of O with Zr and Nb prior to the reaction at the $Nb_3Sn$/Nb heterophase interfaces. Notably, O is segregated at Nb grain boundaries, which leads to a higher concentration of Zr-O and Nb-O dimers at Nb GBs. There is, however, no detectable nucleation of Zr oxide nanoprecipitates in the Nb volume.
- The nucleation of Zr oxide nanoprecipitates occurs actively at $Nb_3Sn$/Nb interfaces, driven by a substantial reduction in the solubilities of Zr and O in $Nb_3Sn$ compared to in Nb. The presence of Nb/$Nb_3Sn$ heterophase interfaces and $Nb_3Sn$ GBs play a crucial role in providing nucleation sites and creating a favorable environment for nucleation of Zr oxide nanoprecipitates.
- The internally oxidized $Nb_3Sn$ wires reacted at 625 °C/740 h exhibit a high number density ($N_v$) of Zr oxide nanoprecipitates (~$10^{23}$ $m^{-3}$) with an average diameter of less than ~10 nm.
- A high number density ($N_v$) of Zr oxide nanoprecipitates effectively hinders grain growth within $Nb_3Sn$ (with average grain diameters less than ~100 nm) for both $Nb_3Sn$ samples reacted at 625 °C and 700 °C.

- The homogeneous nucleation of Zr oxide nanoprecipitates in the $Nb_3Sn$ matrix is analyzed using classical nucleation theory (CNT) and first-principles calculations. According to this analysis, the critical radius ($r^*$) and net reversible work ($W(r)$) for the formation of a spherical Zr oxide nanoprecipitate is estimated: at 625 $^{o}$C, the estimated values are $r^*$=1.2 ±0.1 nm and $W(r^*) = 3.71 \times 10^{-21}$ J, while at 700 $^{o}$C, the values are greater, $r^*$=1.4 ±0.1 nm and $W(r^*) = 4.43 \times 10^{-21}$ J.
- However, heterogeneous nucleation of Zr oxide precipitates is widely observed at the Nb/$Nb_3Sn$ interface and GBs of $Nb_3Sn$ in our APT analysis, and it indicates that actual values of $W(r^*)$ and $r^*$ in the current $Nb_3Sn$ samples are smaller than the calculated values assuming homogeneous nucleation in **Table 1**.
- First-principles calculations and CNT concerning the nucleation of Zr oxide nanoprecipitates demonstrate that low temperatures result in a smaller energy barrier for nucleation of Zr oxide nanoprecipitates, leading to a higher number density of Zr oxide precipitates in $Nb_3Sn$. At low temperatures, diffusion of Zr and O is limited, which impedes the kinetic process of nucleation of Zr oxide nanoprecipitates.
- Our research provides valuable implications for improving the properties of internally oxidized $Nb_3Sn$ wires with a high-number density of Zr oxide nanoprecipitates intended for high-field magnet applications. This framework can also be applied to $Nb_3Sn$ wires with other different types of pinning centers, such as Hf oxides nanoprecipitates.

**Acknowledgments**

This research is supported by the United States Department of Energy, Offices of High Energy. Fermilab is operated by the Fermi Research Alliance LLC under Contract No. DE-AC02-07CH11359 with the United States Department of Energy. We thank Dr. Xuan Peng of Hyper

Tech Research Inc. for providing the internal-oxidation $Nb_3Sn$ wire for this work. This work made use of the EPIC facilities of Northwestern University's NU*ANCE* Center, which received support from the Soft and Hybrid Nanotechnology Experimental (SHyNE) Resource (NSF ECCS-1542205 and ECCS-2025633); the MRSEC program (NSF DMR-1720139 and DMR-2308691) at the Materials Research Center; the International Institute for Nanotechnology (IIN); the Keck Foundation; and the State of Illinois, through the IIN. Atom-probe tomography was performed at the Northwestern University Center for Atom-Probe Tomography (NUCAPT). The LEAP tomograph at NUCAPT was purchased and upgraded with grants from the NSF-MRI (DMR-0420532) and ONR-DURIP (N00014-0400798, N00014-0610539, N00014-0910781, N00014-1712870) programs. NUCAPT received support from the MRSEC program (NSF DMR-1720139 and DMR-2308691) of the Materials Research Center, the SHyNE Resource (NSF ECCS-1542205 and ECCS-2025633), and the Initiative for Sustainability and Energy (ISEN) at Northwestern University.

# Supplementary material

# Unveiling the nucleation and growth of Zr oxide precipitates in internally oxidized $Nb_3Sn$ superconductors

Jaeyel Lee[1,2], Zugang Mao[2], Dieter Isheim[2,3], David N Seidman[2,3], Xingchen Xu[1]

*[1]Fermi National Accelerator Laboratory, Batavia, IL, 60510, USA*

*[2]Department of Materials Science and Engineering, Northwestern University, Evanston, IL, 60208, USA*

*[3]Northwestern University Center for Atom-Probe Tomography (NUCAPT), Evanston, IL, 60208, USA*

**Table S1**. List of the formation enthalpy per atom of different oxide phases: negative signs indicate exothermic reactions ($ZrO_2$, NbO, SnO) [1] and interaction parameters ($\Omega$) between solute atom pairs (Zr-Nb [2], Zr-Sn [3], Nb-Sn [4], Zr-Cu [5])

| Formation Gibbs free energy at 0 K | | Interaction parameters | | |
|---|---|---|---|---|
| | $\Delta G$ (eV/atom) | | $\Omega$ (kJ/mol) | definition |
| $ZrO_2$ | -3.81 | Zr-Nb | 18.4 | $L^{0,bcc}_{Zr-Nb}$ |
| NbO | -2.18 | Zr-Sn | -8.8 | $1/9\ \Delta H_f\ (Zr_5Sn_4)$ |
| SnO | -1.48 | Nb-Sn | -4.0 | $1/4\ \Delta H_f\ (Nb_3Sn)$ |
| | | Zr-Cu | -7.4 | $L^{0,bcc}_{Zr-Cu}$ |

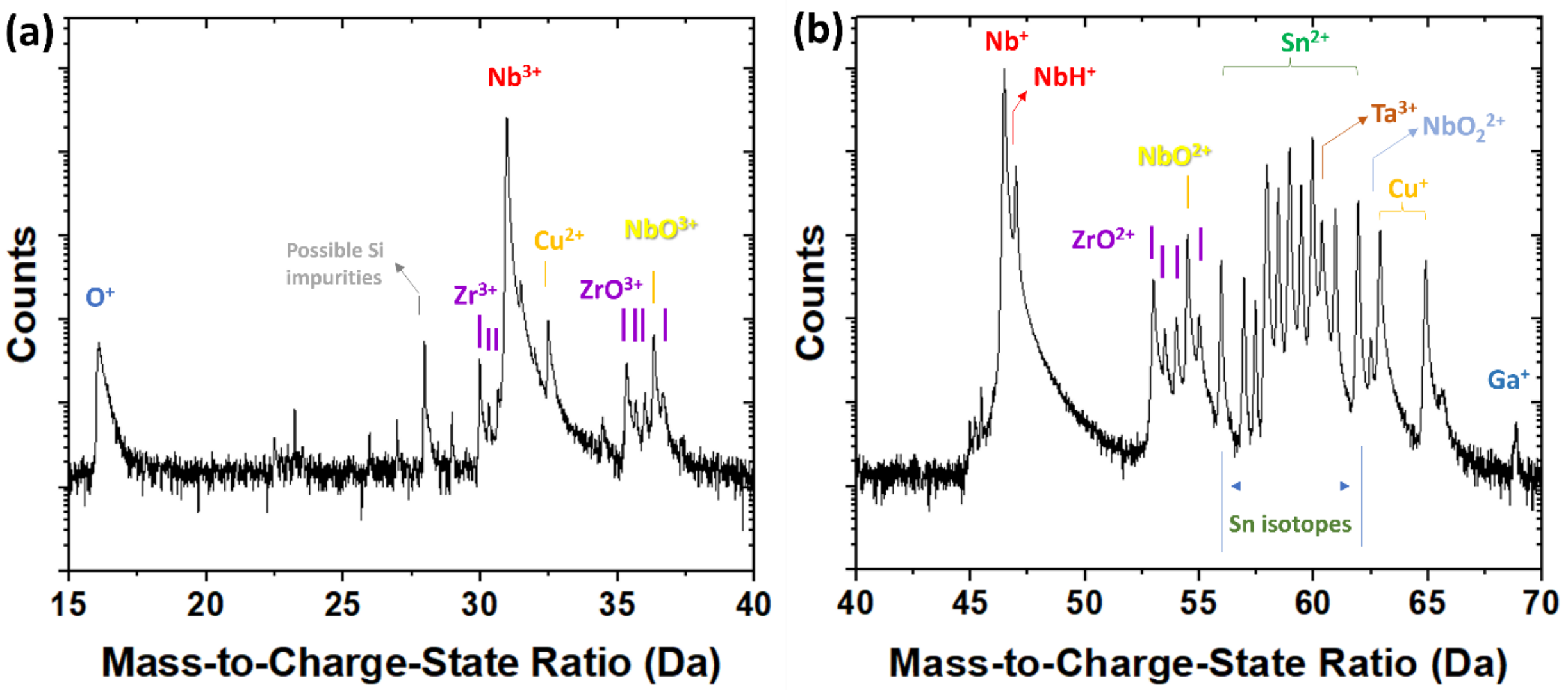

**Figure S1**. (a,b) Mass spectra and identification of peaks in the mass spectra for Nb, Sn, Zr-O, Nb-O, and Cu.

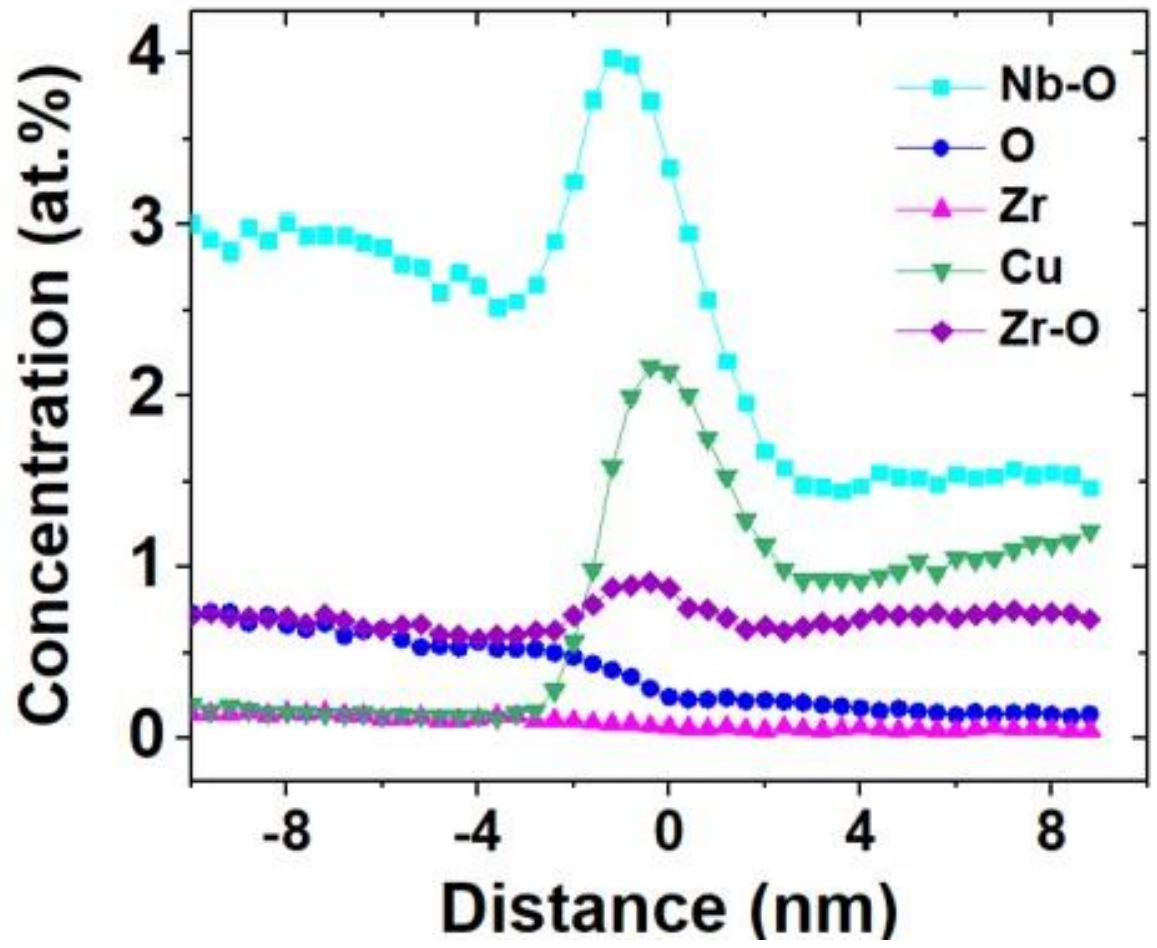

**Figure S2**. Ionic concentration profiles of Nb-O and Zr-O molecular ions with atomic Zr, O, and Cu profiles across the interface, showing a higher concentration of Nb-O and Zr-O at the interface, indicative of segregation and potential formation of nucleation precursors for Zr oxide precipitates at the Nb/$Nb_3Sn$ interface. Cu is also segregated at the $Nb_3Sn$/Nb interface, probably due to enhanced Cu diffusion along the interface.

### S.1 APT analyses of Zr oxide nanoprecipitates in $Nb_3Sn$ layers annealed at 700 °C

Additional APT analyses were performed on the $Nb_3Sn$ sample heat-treated at 700 °C to understand the effect of a higher heat-treatment temperature. Preliminary APT analyses of Zr oxide precipitates in $Nb_3Sn$ heat-treated at 700 °C were reported in ref. [9]. **Fig. S3** displays a 3-D reconstruction of a nanotip of the $Nb_3Sn$ sample, at 5 μm from the Nb/$Nb_3Sn$ interface. The chemical composition of Zr oxide (25.7 ±2.5 Zr, 36.5 ±2.8 O, 21.1 ±2.3 Nb, 13.2 ±1.9 Sn, 1.5 ±0.1 Cu, all in at.%) is closer to stoichiometric $ZrO_2$ than the Zr oxide in the sample annealed at 625 °C at the same distance (5 μm) from the Nb/$Nb_3Sn$ interface. Still, it exhibits partitioning of Nb and Sn to Zr oxide nanoprecipitates, indicating that the equilibrium composition of $ZrO_2$ is not achieved.

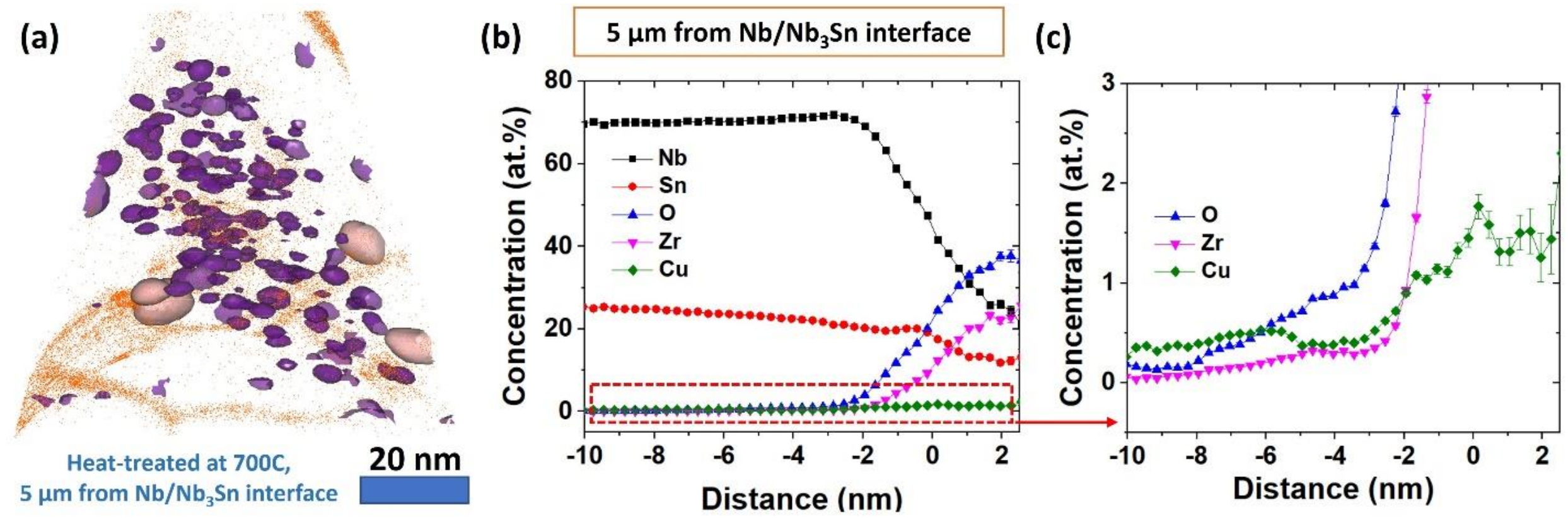

**Figure S3**. (a) 3-D reconstruction of a $Nb_3Sn$ nanotip with Zr oxide nanoprecipitates for a sample aged at 700 °C, which is ~5 μm away from a Nb/$Nb_3Sn$ interface. (b) This plot displays concentration (at.%) profiles across interfaces between four representative Zr oxide nanoprecipitates and $Nb_3Sn$ using the proximity histogram methodology in IVAS. (c) The ordinate of this plot is in concentration values ranging from 0 to 3 at. %, which displays the atomic concentrations of Cu in Zr oxide nanoprecipitates in detail. This plot reveals large concentrations of Zr (25.7±2.5 at.%) and O (36.5±2.8 at.%), plus partitioning of Nb (21.1±2.3 at.%), Sn (13.2 ±1.9 at.%), and Cu (1.5±0.2 at.%) into the cores of four large selected Zr oxide precipitates.

### S.2 Evolution of Zr oxide precipitates in $Nb_3Sn$ layers

Thermodynamic estimations of the Nb-Sn-Zr-O system for the formation energies of each phase, **Table S1**, predict that the equilibrium system is a mixture of three phases, $Nb_3Sn + ZrO_2 + NbO$, for the given alloy composition, Nb-Sn-Zr-O (25.0 Sn, 0.7 Zr, 2.1 O, all in at.%) at 625 °C. There is excess oxygen for the formation of $ZrO_2$ in the $Nb_3Sn$ matrix, and the excess oxygen atoms form NbO, which has the second-largest negative heat of formation of an oxide phase close to $ZrO_2$, in addition to $ZrO_2$ in the $Nb_3Sn$ matrix. Indeed, we observed complex compositions of Zr oxide nanoprecipitates, including Nb and Sn in the Zr oxide precipitates. Considering that the specimen is in a nonequilibrium state and the low-temperature range (625-700 °C), which is ~0.4 $T_m$ of $Nb_3Sn$, it is not that surprising that Zr oxide has a complicated composition and a non-equilibrium cubic structure. It has been reported that Zr oxides have a possible non-equilibrium phase, $ZrO_x$ ($0<x<2$) as an intermediate state [6, 7]. Furthermore, there is evidence of complicated compositions of oxide phases, $Nb_{1-x}Zr_xO_2$ or $ZrO_{1-y}Nb_yO_x$ in the Nb-Zr-O ternary system [8].

The chemical evolution of Zr oxide precipitates is influenced by several thermodynamic and kinetic factors, such as chemical interactions among solute species, which are discussed in the **Supplementary Information, Table S1** and **Fig. S5**. **Fig. S4(a)** reveals the compositional evolution of Zr oxide precipitates in $Nb_3Sn$ layers. Several factors play roles in the evolution of the Zr oxide precipitates: (i) slow diffusion of Zr in $Nb_3Sn$; (ii) the existence of excess oxygen in $Nb_3Sn$; and (iii) chemical interactions among Nb, Sn, Zr, and O atoms in the Zr oxide precipitates, **Table S1** [3, 9, 10]. Firstly, diffusion of Zr in $Nb_3Sn$ is anticipated to be slow similar to the slow diffusion of Zr in Nb, as $Nb_3Sn$ is an ordered structure, where the diffusion path of Zr is limited to specific sites [11]. This may cause partitioning of Nb and Sn to Zr oxides, prior to forming the equilibrium $ZrO_2$ phase. Secondly, the existence of excess oxygen in $Nb_3Sn$ may cause Nb to react with oxygen and be incorporated in Zr oxide nanoprecipitates. Lastly, the chemical interactions among solute atoms play roles in the compositional evolution of Zr oxide nanoprecipitates. Niobium has repulsive interactions with Zr [2] and Sn has attractive interactions with Zr [3], **Table S1**, which is reflected in the compositional evolution of Zr oxide precipitates, **Fig. S4(a)**. The decrease of the Nb concentration is faster than that of Sn in Zr oxides during temporal evolution. The partitioning of a small concentration of Cu in Zr oxide precipitates, 1.5±0.1 at.% at 700 ºC, is rationalized by attractive interactions between Zr and Cu [5].

Zirconium oxides have different crystal structures: (i) monoclinic below 1170 ºC; and (ii) tetragonal and cubic structures at higher temperatures [12]. Also, possible non-equilibrium phases, ZrO and $ZrO_{2-x}$ (0<x<2) are observed employing APT, TEM experiments, and first-principles calculations [6, 7, 13]. In our system, where Zr oxides are embedded in the $Nb_3Sn$ matrix, an HR-STEM image, **Fig. 5**, suggests that the structure of some Zr oxides is possibly a cubic structure with a lattice parameter of ~5.4 Å. This result is different from the equilibrium phase of Zr oxide, monoclinic $ZrO_2$ below 1170 ºC [12], which is rationalized by the effect of the cubic structure of the $Nb_3Sn$ matrix.

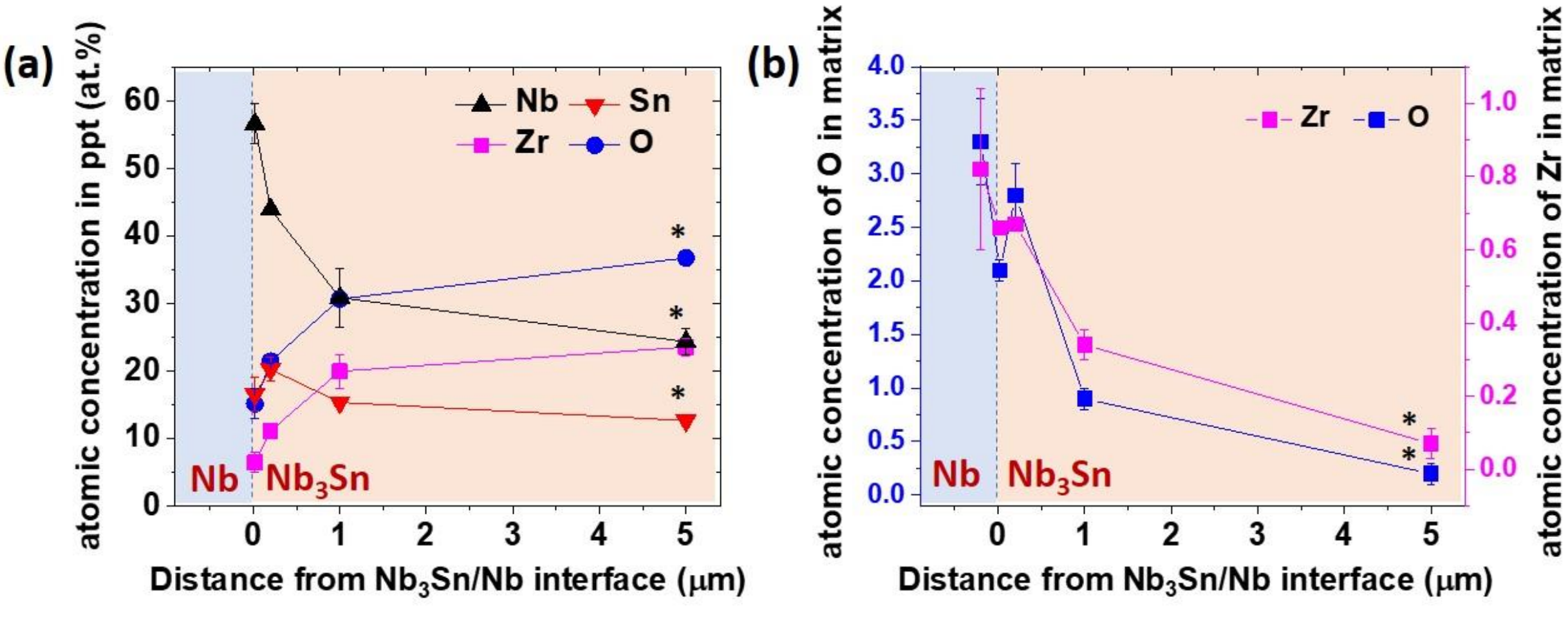

**Figure S4**. (a) Atomic concentrations of Zr, O, Nb, and Sn in Zr oxide nanoprecipitates in dependence from distance from the Nb/$Nb_3Sn$ interface; and (b) atomic concentrations of Zr and O in Nb and the $Nb_3Sn$ matrix. $Nb_3Sn$ wire is prepared at 625 ºC and the atomic concentrations of each element are determined employing APT analyses. Zirconium oxide nanoprecipitates contain significant Nb concentrations at early stage, that is, close to the Nb/$Nb_3Sn$ interface, and gradually approach the equilibrium composition, $ZrO_2$ at larger distances from the Nb/$Nb_3Sn$ interface. The data at 5 μm, denoted by '*', is extracted from a $Nb_3Sn$ wire prepared at 700 ºC, which is close to the equilibrium concentration due to the higher aging temperature. For Nb and the $Nb_3Sn$ matrix,

the atomic concentrations of Zr and O are high initially in the Nb matrix and gradually decrease in the $Nb_3Sn$ matrix.

In case a significant local magnification effect occurs in atom-probe tomography between the nanoprecipitates and the $Nb_3Sn$ matrix, it causes trajectory overlaps of the evaporating ions between the nanoprecipitates and the Nb matrix along a horizontal direction in the 3D APT reconstruction. In order to investigate that question, we compared 1-D compositional analyses, using IVAS, across Zr oxide nanoprecipitates in two different directions in the 3D APT reconstruction: one in a horizontal direction and another one in a vertical direction to analyze a possible effect of local magnification and ion trajectory overlaps and there is no noticeable difference, indicating that the effect of local magnification is insignificant, **Fig. S5**. APT dataset of Zr oxide precipitates in $Nb_3Sn$ heat-treated at 700 °C were also reported in ref. [9] and **Fig. S3**.

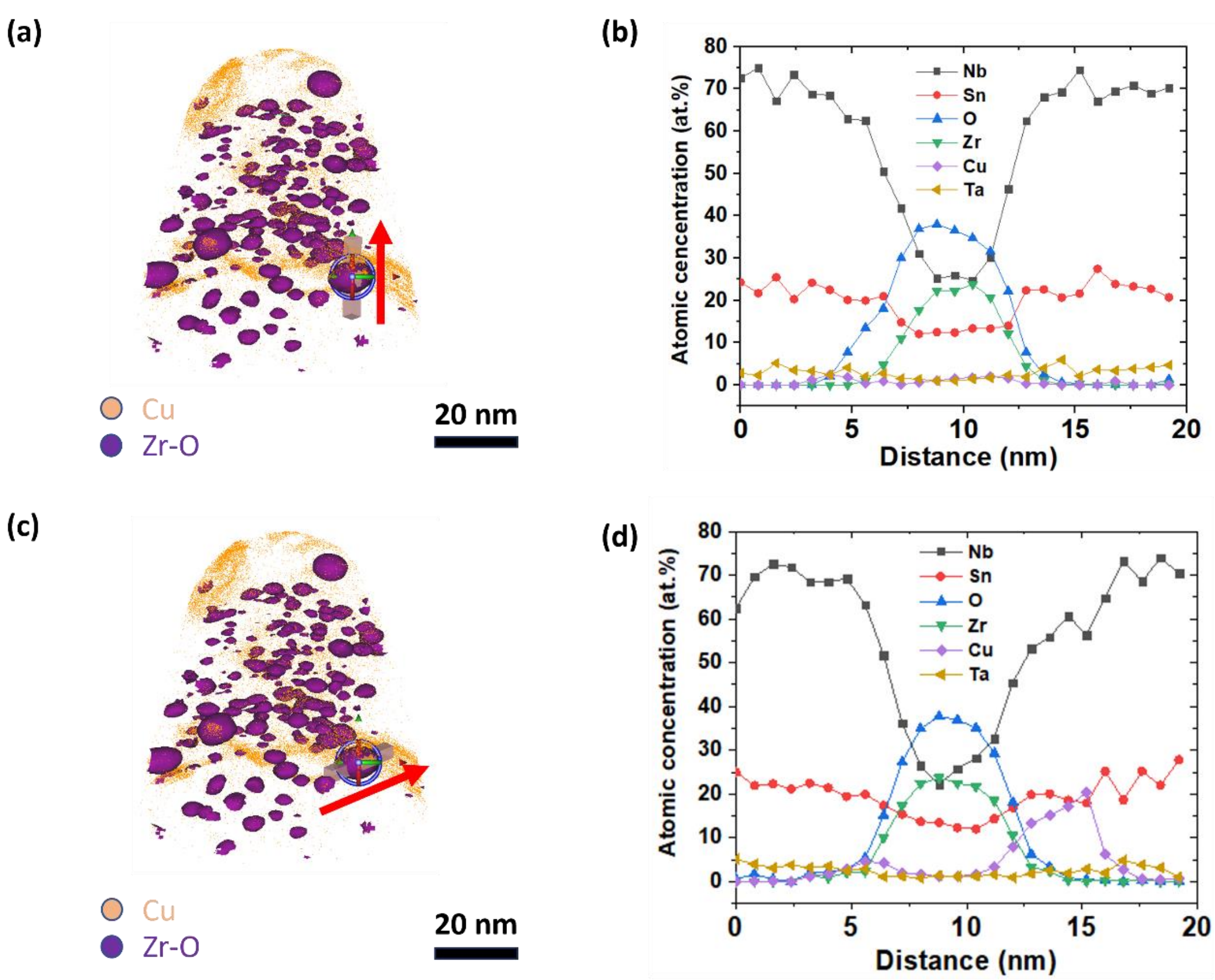

**Figure S5** 1-D concentration profiles across a Zr oxide nanoprecipitate in an internally oxidized $Nb_3Sn$ wire aged at 700 °C for two different measurement directions: (a) vertical red arrow next to the probe cylinder indicating the direction of the profile and (b) the concentration profile along that direction; and (c) horizontal red arrow indicating the direction of the concentration profile displayed in (d). There aren't significant differences in the chemical compositions of the Zr oxide precipitate observed from the 1D-concentration profiles for the two different directions studied, (a) and (c), which indicates that a local magnification effect is insignificant for Zr oxide precipitates/$Nb_3Sn$ heterophase interfaces and does not affect the concentration measured in the center of the precipitates significantly. We also note that the concentration of Ta in Zr oxide precipitates decreased following the trend of the concentration of Nb in the Zr oxide precipitates. The APT dataset of Zr oxide precipitates in Fig. S4 was also used in ref. [9] and **Fig. S3**.

### S.3 First-principles calculations of the chemical formation energy using the Helmholtz free energy change during nucleation ($\Delta F_{chem}$): The elastic strain energy ($\Delta E_{elastic}$) and interfacial energy of the $ZrO_2$/$Nb_3Sn$ matrix interface ($\sigma$)

Firstly, we estimate the chemical Helmholtz free energy change during the nucleation ($\Delta F_{chem}$) of $ZrO_2$ nanoprecipitates. To understand the initial mixed state of $Nb_3Sn$ and Zr-O, the Zr substitutional $Nb_3Sn$ structures were modeled by allowing Zr to substitute at either the Sn or Nb sublattice sites in the 2×2×2 $Nb_3Sn$ superlattices, which are fully relaxed. The substitutional energies in $Nb_3Sn$ are calculated according to the following equations:

$$E_{Zr\rightarrow Sn} = (E^{TOT}_{Nb_3(Sn_{1-x}Zr_x)} + \mu_{Sn}) - (E^{TOT}_{Nb_3Sn} + \mu_{Zr}) \quad (4)$$

$$E_{Zr\rightarrow Nb} = (E^{TOT}_{(Nb_{1-y}Zr_y)_3Sn} + \mu_{Nb}) - (E^{TOT}_{Nb_3Sn} + \mu_{Zr}) \quad (5)$$

where $\mu_i$ is the chemical potential of Nb or Sn, $E^{TOT}_{Nb_3Sn}$ is the total energy of bulk $Nb_3Sn$, $E^{TOT}_{Nb_3(Sn_{1-x}Zr_x)}$ is the total energy of $Nb_3Sn$ with Zr substituting in the Sn sublattice, and $E^{TOT}_{(Nb_{1-y}Zr_y)_3Sn}$ is the total energy of $Nb_3Sn$ with Zr substituting in the Nb sublattice. First-principles results show that $E_{Zr\rightarrow Nb}$ (0.124 eV/atom) is significantly smaller than $E_{Zr\rightarrow Sn}$ (0.475 eV/atom), therefore Zr prefers Nb sublattice sites instead of Sn and we employ $(Nb_{3(n-1)}Zr_3)O_6Sn_n$ as representing the solid solution.

The formation energy of cubic ZrO2 precipitate ($\Delta F_{chem,ZrO2}$) is estimated as from the following equation:

$$\Delta F_{chem}=[(n-1)\Delta F(Nb_3Sn)+3\Delta F\left(cubic_{ZrO_2}\right)+\Delta F_{Sn}]-[\Delta F\left((Nb_{3(n-1)}Zr_3)O_6Sn_n\right)+3\Delta F_{Nb}]\approx\left\{\left[n\Delta H(Nb_3Sn)+3\Delta H\left(cubic_{ZrO_2}\right)\right]-[\Delta H\left((Nb_{3(n-1)}Zr_3)O_6Sn_n\right)]\right\}$$

$$-T\left\{\left[n\Delta S(Nb_3Sn)+3\Delta S\left(cubic_{ZrO_2}\right)\right]-[\Delta S\left((Nb_{3(n-1)}Zr_3)O_6Sn_n\right)]\right\}$$

(6)

Where $\Delta H\left((Nb_{3(n-1)}Zr_3)O_6Sn_n\right)$ is the total enthalpy of the $Nb_3Sn$ solid solution with Zr substituting for Nb and O in interstitial sites: $\Delta H(Nb_3Sn)$ is the bulk formation enthalpy of $Nb_3Sn$, and $\Delta H(cubic_ZrO_2)$ is the bulk formation enthalpy of $ZrO_2$. We take the enthalpy, H, to be equal to the internal energy, E, because the pressure-volume term in H is negligible compared to E in the solid state. The entropy terms $\Delta S$ in Eq. 6 are from the phonon vibrational entropies. In a dilute $(Nb_{3(n-1)}Zr_3)O_6Sn_n$ solution, we don't need to consider its contribution because the configurational entropy is smaller than the vibrational entropy. The calculated phonon vibrational entropies for $Nb_3Sn$, $ZrO_2$, and $(Nb_{3(n-1)}Zr_3)O_6Sn_n$ solid-solution as a function of temperature are explained and plotted in **Fig. S6**.

Secondly, we calculate the interfacial energy and elastic strain energy of the $ZrO_2$/$Nb_3Sn$ interface using first-principles calculations. In our first-principles calculations, we find that monoclinic $ZrO_2$ has the lowest formation energy and cubic $ZrO_2$ has the highest formation energy. Experimentally, we observed the $ZrO_2$ (110) /$Nb_3Sn$ (110) interfacial structure with a misfit of 2.6%. The crystal structure of $ZrO_2$ is close to a cubic structure in a TEM image and its diffraction pattern. We adopted the cubic structure for all our interfacial calculations. The bulk energies of $Nb_3Sn$ and $ZrO_2$, $E^{tot}_{\mathrm{ZrO_2}}$ and $E^{tot}_{\mathrm{Nb_3Sn}}$, are obtained from the formation energy of each structure, which are relaxed to their own equilibrium state. We calculated the interfacial energy using first-principles calculations employing the following equation:

$$\sigma_{\mathrm{ZrO_2/Nb_3Sn}} = \frac{1}{2A}\left[E^{tot}_{\mathrm{ZrO_2/Nb_3Sn}} - (E^{tot}_{\mathrm{ZrO_2}} + E^{tot}_{\mathrm{Nb_3Sn}})\right] \tag{7}$$

Where $A$ is the interfacial area, $E^{tot}_{\mathrm{ZrO_2/Nb_3Sn}}$ is the total internal energy of the relaxed $ZrO_2/Nb_3Sn$ system containing an interface. The resulting calculated $ZrO_2$ (110) /$Nb_3Sn$ (110) interfacial internal energy at 0 K is 325 $mJ/m^2$.

Thirdly, the strain energy of the $ZrO_2/Nb_3Sn$ interface is calculated using first-principles calculations. The strain is generated by a volume distortion and calculated according to volume matching of $Nb_3Sn$ to $ZrO_2$, due to the strong Zr-O bonding. The strain energy was calculated from the following equation [14]:

$$E^{hkl}_{ZrO_2/Nb_3Sn\ strain} = (E^{bulk}_{constrained} - E^{bulk}_{unconstrained}) / n_{Nb_3Sn} \tag{8}$$

Where $E^{bulk}_{constrained}$ is the total energy of constrained bulk $Nb_3Sn$ with an (*hkl*) plane, matching to the $ZrO_2$ (*h'k'l'*) interface plane, $E^{bulk}_{unconstrained}$ is the total energy of unconstrained bulk $Nb_3Sn$, and $n_{Nb_3Sn}$ is the total number of atoms in bulk $Nb_3Sn$. The calculated strain energy at 0 K, $E^{hkl}_{ZrO_2/Nb_3Sn\ strain}$, is 0.194 eV/atom.

To calculate the phonon dispersions for $Nb_3Sn$, $ZrO_2$, and the $(Nb_{3(n-1)}Zr_3)O_6Sn_n$ solid-solution, we used the frozen-phonon method for 3×3×3 supercells for bulk supercells. Using this methodology, we calculated a series of total energy differences between atomic configurations with and without a single displaced atom. The harmonic approximation was employed for all our vibrational entropy calculations [15-17].

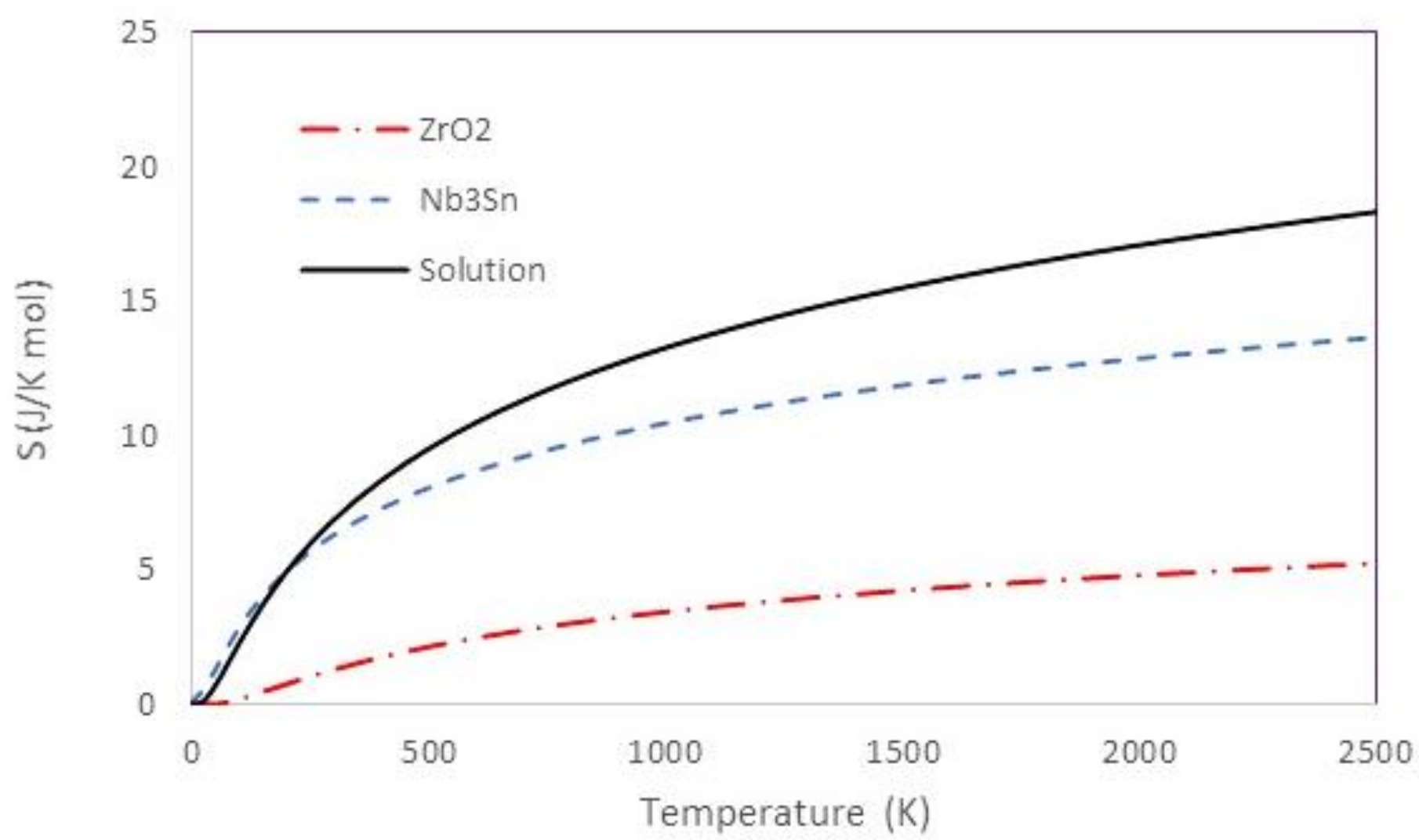

**Figure. S6** The calculated phonon vibrational entropies for $Nb_3Sn$, cubic $ZrO_2$, and a $(Nb_{3(n-1)}Zr_3)O_6Sn_n$ solid-solution as a function of temperature.